\documentclass[iop]{emulateapj} 
\usepackage[usenames,dvipsnames]{color}
\usepackage{xspace}

\usepackage{apjfonts}
\slugcomment{Accepted for publication in ApJ}

\shorttitle{FIR CO Evolution}
\shortauthors{Manoj et al.}

\newcommand{\ee}[1]{${} \times 10^{#1}$}
\newcommand{\eten}[1]{$10^{#1}$}

\newcommand{\kms}{\textrm{km~s}\ensuremath{^{-1}}}	
\newcommand{\percc}{\ensuremath{\textrm{cm}^{-3}}\xspace}

\newcommand{\lsun}{\ensuremath{L_{\odot}}\xspace}			
\newcommand{\lbol}{\ensuremath{L_{\mathrm{bol}}}\xspace}	 
\newcommand{\lphot}{\ensuremath{L_{\mathrm{phot}}}\xspace}	 
\newcommand{\lacc}{\ensuremath{L_{\mathrm{acc}}}\xspace}	 
\newcommand{\tbol}{\ensuremath{T_{\mathrm{bol}}}\xspace}	
\newcommand{\macc}{\ensuremath{\dot{M}_{\mathrm {acc}}}\xspace}	
\newcommand{\maccav}{\ensuremath{\langle \dot{M}_{\mathrm {acc}}\rangle}\xspace}	
\newcommand{\mout}{\ensuremath{\dot{M}_{\mathrm{out}}}\xspace}	
\newcommand{\moutav}{\ensuremath{\langle \dot{M}_{\mathrm{out}}\rangle}\xspace}	
\newcommand{\vout}{\ensuremath{{v}_{\mathrm{out}}}\xspace}

\newcommand{\fsmm}{{$L_{\mathrm{smm}}/L_{\mathrm{bol}}$}\xspace} 
\newcommand{\lsmm}{{$L_{smm}$}\xspace} 
\newcommand{\lbolsmm}{{$L_{\mathrm {bol}}/L_{\mathrm {smm}}$}\xspace} 

\newcommand{\spitzer}{{\it Spitzer}\xspace}

\newcommand{\herschel}{{\it Herschel}\xspace}

\newcommand\submm{submillimeter\xspace}
\newcommand\smm{submillimeter\xspace}
\newcommand{\fir}{far-IR\xspace}

\newcommand{\jj}[2]{{$J = #1\rightarrow#2$}\xspace} 

\newcommand{\menv}{{$M_{\rm env}$}\xspace}
\newcommand{\lco}{{$L^{\rm fir}_{\rm CO}$}\xspace} 
\newcommand{\lcolow}{{$L_{\rm {CO} (J=14-13)}$}\xspace} 
\newcommand{\lcomm}{{$L^{\rm (sub)mm}_{\rm CO}$}\xspace} 
\newcommand{\fco}{{$F_{\rm CO}$}\xspace} 
\newcommand{\lmech}{{$L_{\rm mech}$}\xspace}
\newcommand{\nhh}{{$n\mathrm{(H_2)}$}\xspace}

\begin{document} 

\title{The evolution of far-infrared CO emission from protostars}
\author{P. Manoj\altaffilmark{1}, J. D. Green\altaffilmark{2}, S. T. Megeath\altaffilmark{3}, N. J. Evans II \altaffilmark{4}, A. M. Stutz\altaffilmark{5}, J. J. Tobin\altaffilmark{6}, D. M. Watson\altaffilmark{7}, W. J. Fischer\altaffilmark{8}, E. Furlan\altaffilmark{9}, T. Henning\altaffilmark{5}}

\altaffiltext{1}{Tata Institute of Fundamental Research, Homi Bhabha Rd, Mumbai 400 005}
\email{manoj.puravankara@tifr.res.in}

\altaffiltext{2}{Space Telescope Science Institute, Baltimore, MD, USA}

\altaffiltext{3}{Department of Physics and Astronomy, University of Toledo, 2801 West Bancroft Street, OH 43606, USA}

\altaffiltext{4}{The University of Texas at Austin, Department of Astronomy, 2515 Speedway, Stop C1400, Austin, TX 78712-1205, USA}

\altaffiltext{5}{Max-Planck-Institute for Astronomy, K${\rm \ddot{o}}$nigstuhl 17, 69117 Heidelberg, Germany}

\altaffiltext{6}{Leiden Observatory, Leiden University, P.O. Box 9513, 2300-RA Leiden, The Netherlands}

\altaffiltext{7}{Department of Physics and Astronomy, University of Rochester, Rochester, NY 14627, USA}

\altaffiltext{8}{NASA Goddard Space Flight Center, Greenbelt, MD, USA}

\altaffiltext{9}{Infrared Processing and Analysis Center, California Institute of Technology, 770 S. Wilson Ave., Pasadena, CA 91125, USA}

\begin{abstract}
We investigate the evolution of \fir CO emission from protostars observed with \herschel/PACS for 50 sources from the combined sample of  HOPS and DIGIT \herschel key programs. From the uniformly sampled spectral energy distributions, whose peaks are well sampled, we computed the \lbol, \tbol and \lbolsmm for these sources to search for correlations between \fir CO emission and protostellar properties. We find a strong and tight correlation between \fir CO luminosity (\lco) and the bolometric luminosity (\lbol) of the protostars with \lco~$\propto$~\lbol~$^{0.7}$.  We, however, do not find a strong correlation between \lco and protostellar evolutionary indicators, \tbol and \lbolsmm.  FIR~CO emission from protostars traces the currently shocked gas by jets/outflows, and \fir CO luminosity, \lco, is proportional to the instantaneous mass loss rate, \mout. The correlation between \lco and \lbol, then, is  indicative of instantaneous \mout  tracking instantaneous \macc. The lack of correlation between \lco and evolutionary indicators \tbol and \lbolsmm suggests that \mout and, therefore,  \macc do not show any clear evolutionary trend. These results are consistent with mass accretion/ejection in protostars being episodic. Taken together with the previous finding that the time-averaged mass ejection/accretion rate declines during the protostellar phase \citep[e.g.][]{bontemps96, curtis10}, our results suggest that the instantaneous accretion/ejection rate of protostars is highly time variable and episodic, but the amplitude and/or frequency of this variability decreases with time such that the time averaged accretion/ejection rate declines with system age.
 \end{abstract}


\section{Introduction} \label{sec1}

Mass accretion in young stellar objects is thought to be highly time variable and episodic \citep[e.g.][]{ken90, hartken96, hart09, evans09, dunham12}.  Outbursts of varying intensities and frequencies have been observed in several young stars \citep[e.g][]{herbig77,  kenyon95, green06, ra10, fischer12, green13b, audard14, safron15}. However, a detailed picture of the time evolution of mass accretion from early protostellar phase to late pre-main sequence  phase is still missing. Most of the commonly used direct observational tracers of mass accretion fall at wavelengths $\la$~2~\micron~\citep[e.g.][]{cg98, mch98, muz98, muzerolle01}. While these tracers are used extensively to study the accretion history in pre-main sequence stars \citep[e.g.][]{ghbc98, hart98, calvet04, hh08, muz04}, they are difficult to observe in protostars which are deeply embedded in their natal core as their shorter wavelength emission is heavily extinguished.

Mass accretion in protostars is thought to be associated with mass ejection. Jets and outflows from embedded protostars are more readily accessible to observations than the direct accretion tracers, particularly at  \fir~and (sub)mm~wavelengths. Observations of protostellar jets and outflows at these wavelengths provide important diagnostics for the energetics of mass ejection and mass loss rates from protostars \citep[e.g.][]{bt99, richer00, tobin16, nisini15, watson85, hollen89}. Moreover, theoretical models of mass ejection mechanisms from protostars predict a linear relation between mass loss rate from protostars, \mout, and mass accretion rate, \macc,  onto the protostar \citep{shu94, ns94, pp92, wk93, mp05, mp08}.   Thus, observed properties of mass ejection can be used to study the mass accretion history in protostars.

While jets and outflows from protostars have been studied using several different tracers and at various wavelengths \citep[e.g.][]{frank14, bally07}, the observational tool that is most often used for the largest sample of protostars are the low-$J$ ($J_{up}~\leq$ 3) CO lines at (sub-)mm wavelengths which trace the ambient molecular  gas swept up and accelerated by the protostellar jets  \citep[e.g.][]{richer00,  bt99, arce07, hatchell07, takahashi08, curtis10, dunham14b, plunkett15b}. These observations of molecular outflows from protostars have shown that the time-averaged flow energetic parameters, viz., the mechanical luminosity (\lmech) and the momentum flux or outflow force (\fco), are tightly correlated with the bolometric luminosity of the protostar, \lbol~\citep{rodri82, bl83, lada85, snell87, cb92, bontemps96, wu04,  hatchell07, takahashi08, curtis10}.  A few of these studies also found an evolutionary trend in the  outflow force (\fco) with protostellar age \citep{bontemps96, curtis10}. Younger Class~0 sources are found to have more powerful outflows than the more evolved Class~I protostars and the outflow power is found to decrease with system age.  This has been interpreted as due to a corresponding steady decline in the mass accretion rate with time during the protostellar phase \citep{bontemps96}.

Emission lines due to the rotational transitions of CO in the \fir~($14~\leq~J_{up}~\leq~45$ ), observed with the PACS instrument onboard the \herschel~space telescope, provide an alternate diagnostic of the jets/outflow properties of protostars. Unlike the low-$J$ CO lines observed at (sub)mm~wavelengths which trace the ambient molecular gas swept up by the jets/outflows, the \fir~CO lines trace the hot gas that is currently being shocked by the jets/outflows from protostars. While the low-$J$ CO lines provide time-averaged energetics of the jets, the \fir~lines provide the instantaneous energetics of the jets. In this paper, we investigate the evolution of jet/outflow properties derived from the \fir~CO line luminosities for a large sample of protostars observed with \herschel/PACS~ as part of the two \herschel~key programs, Herschel Orion Protostar Survey (HOPS) \& Dust, Ice, and Gas in Time (DIGIT). 

\section{The Sample}\label{sec2}

We analysed the combined HOPS  \citep{manoj13} and DIGIT \citep {green13} sample of protostars for which we have \herschel/PACS spectra to search for evolutionary trends in the \fir~CO emission observed towards protostars. The HOPS program was a \fir survey of \spitzer identified protostars in the Orion molecular clouds. The HOPS team obtained and analysed \herschel/PACS photometry of 330 protostars and PACS spectra of 36 protostars \citep{fischer10, stanke10, fischer13, manoj13, stutz13, furlan16, bea16}. A detailed analysis of the  \fir spectra of the 21 brightest sources in the HOPS spectroscopy sample were presented in  \citet{manoj13}. From the \herschel/PACS imaging data, the HOPS team has also identified and characterised 16 new protostars in Orion which were not detected by \spitzer or are too faint at mid-IR wavelengths \citep{stutz13, tobin15}. These sources, which are the reddest, and potentially youngest protostars in the Orion molecular clouds were called  PACS Bright Red sources (PBRs) \citep{stutz13}. \herschel/PACS spectra for 8 PBRs were later obtained by the  HOPS team as part of the \herschel open time program \citep{tobin16}.  The DIGIT program surveyed 94 young stellar objects in different evolutionary stages with \herschel and obtained \fir spectra of 24 Herbig Ae/Be stars, 40 T Tauri stars and 30 protostars in nearby (d $\la$~400~pc) star forming regions \citep{vankemp10b, cieza13, sturm13, Meeus13, Fedele13, dionatos13, green13, green13b, lee14, lee14b, green16}. The \herschel/PACS spectra of 30 protostars in the DIGIT sample were presented in \citet{green13}.

All the 21 HOPS protostars whose \fir spectra were presented in \citet{manoj13} are inlcuded in the current analysis. However, only 21 of the 30 DIGIT sources presented in \citet{green13} are considered here. The sources in the DIGIT sample are on average much closer than the HOPS sources, which complicates the sample in two ways: the sources are slightly extended at various wavelengths (depending on the envelope temperature structure) and some of the source fields (e.g. RCrA, Serpens) are crowded.  The improved data pipeline \citep[see][]{green16} including jitter correction can successfully correct for slightly extended source continuum, but only for the brighter sources, and only when the target is the dominant source in the field of view, within a spaxel of the center. It also assumes that the line emission is distributed identically to the continuum.  The remaining 9 DIGIT sources did not meet these criteria and are excluded; this includes the complicated regions RCrA-IRS5A, 7B, and 7C and Serpens-SMM3/4; the chain of MM sources in the L1448 field; and the off-center IRS 46/44 field.  Also included in our analysis are the 8 PACS Bright Red sources (PBRs) for which \herschel/PACS \fir spectra have been obtained \citep{tobin16}. Our final sample consists of a total of 50 protostars, which are listed in Table~\ref{tab1}.

\begin{figure*}
\includegraphics[scale=0.475]{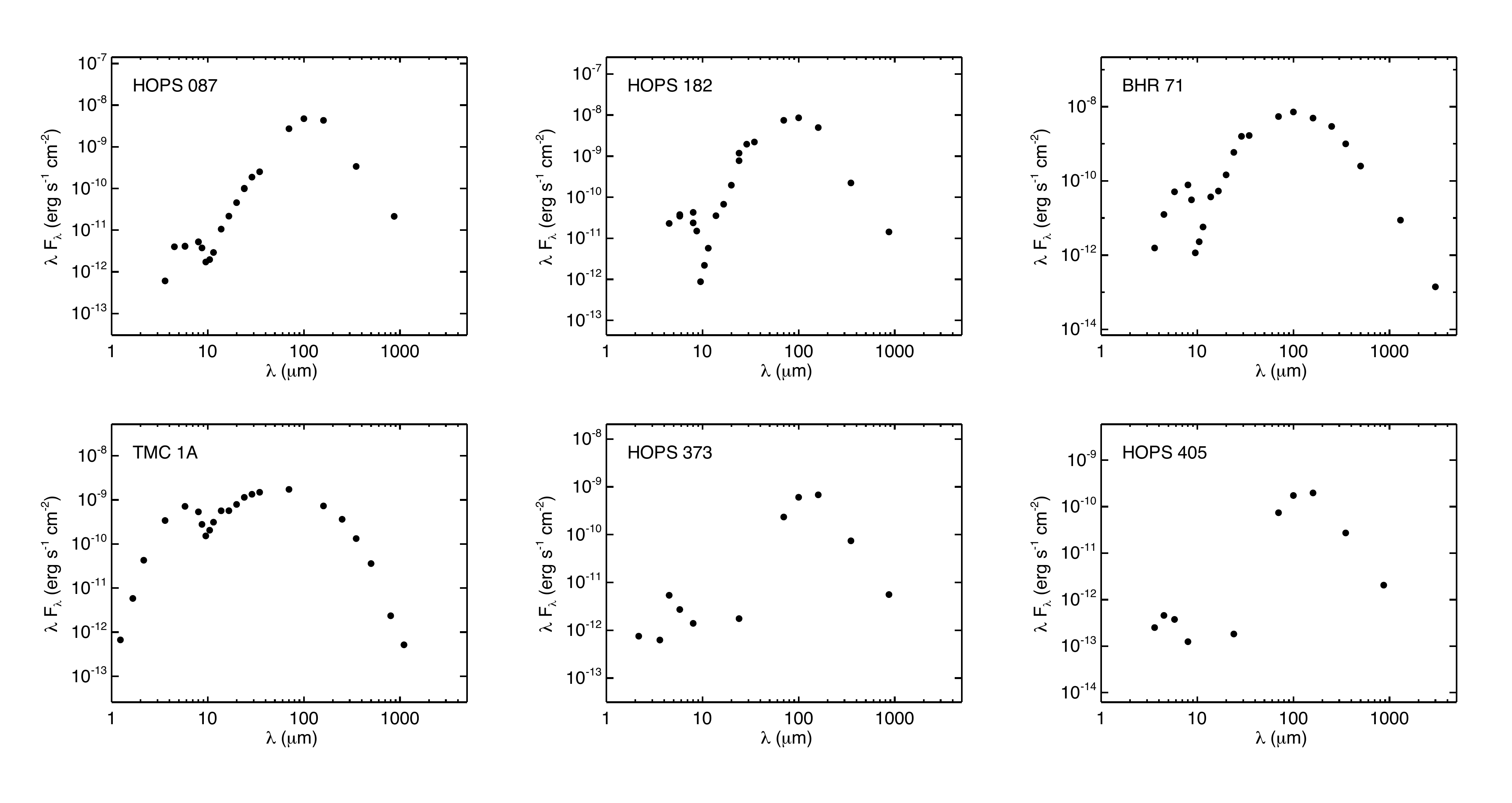}
\caption{ Representative SEDs of six protostars from the sample listed in Table~\ref{tab1}. The near-, mid- and far-IR data points shown are from 2MASS, \spitzer IRAC, MIPS \& IRS and \herschel PACS. \spitzer-IRS spectra have been rebinned to 12 flux values. The submm/mm data for HOPS and PBRs are from APEX \citep{furlan16} and for DIGIT sources from \herschel/SPIRE and ground based submm/mm photometry compiled by \citet{green13} \label{fig0}}
\end{figure*}

\section{Data \& Analysis} \label{sec3}
\subsection{Spectral Energy Distributions}
We first constructed the observed spectral energy distributions (SEDs) of our sample sources in order to estimate the basic protostellar properties. The SEDs of all the sources in our sample are uniformly sampled, from near-IR to the submm/mm. For all sources we have used \herschel/PACS 70, 100 and 160~$\mu$m fluxes to make sure that the peak of the protostellar SED is well sampled. Figure~\ref{fig0} shows a few representative SEDs. Below we provide a detailed description of the photometric data used in the construction of the SEDs of the sources in our sample.

\subsubsection{HOPS} \label{sec311}
 For the HOPS sample, the 1.2 to 870~\micron~photometric data used are from 2MASS, \spitzer IRAC, MIPS and IRS, \herschel PACS and Atacama Pathfinder Experiment (APEX) \citep[][]{manoj13, stutz13, furlan16}. The \spitzer IRS spectra have been rebinned to 12 flux values before integrating the SEDs. The details of the source of the photometry and the apertures used to extract the flux are described in the Appendix of  \citet[][]{manoj13}.  However, there are a few differences in the photometric data used in this analysis and those presented in \citet{manoj13}. The primary one is in the PACS photometry:  for several sources in crowded regions and/or regions where significant extended emission is present, we have updated the 70 and 160~\micron~flux using point spread function (psf) photometry; in addition, for sources which have been observed multiple times in overlapping fields, the averaged value of flux density is quoted \citep[][Ali et al., in preparation]{furlan16}. The PACS 70~\micron~fluxes used in this analysis differ from those presented in \citet{manoj13} by $\sim$~4\%~(median); for 75\% of the  sources the difference is within~11\%. However, for two sources, HOPS 108 and 369, the 70~\micron~flux is lower by more than~60\%.  These sources are in the crowded OMC-2 region which also has significant extended emission and we have used psf photometry to update the aperture photometry reported in \citet{manoj13}. The median change in the PACS 160~\micron~flux is $\sim$~9\%; for 75\% of the sources the difference is within 20\%.  For 11 sources we have updated the 160~\micron~flux using psf photometry. The median difference from those presented in \citet{manoj13} $\sim$~11\% and the maximum difference is in the case of HOPS 91 and 108 which have their 160~\micron~flux lower by $\sim$~45\% compared to that in \citet{manoj13}.   In this analysis, we have  also added PACS 100~\micron~photometry for our sources obtained from the Gould Belt Survey \citep[e.g.][]{andre10}.  The aperture used to extract the the 100~\micron~flux is same as that used for 70~\micron~\citep[for details see][]{furlan16} The 350~\micron~and 870~\micron~data from APEX SABOCA and LABOCA were also re-reduced and re-calibrated \citep{stutz13, stutz15}.  The changes in flux values for most sources are $\sim$~10\%, well within the nominal flux uncertainties of $\sim$~30\%. 
 
%

\subsubsection{DIGIT} \label{sec312}
The photometric data used to construct the SEDs of DIGIT sources have similar wavelength coverage and sampling as that for HOPS sources. The 2MASS and \spitzer-IRAC, MIPS and IRS data were compiled from those presented in \citet[][and references therein]{green13}. As for HOPS sources, the \spitzer IRS spectra have been rebinned to 12 flux values before integrating the SEDs. In addition, we obtained the PACS 70, 100 \& 160~\micron and SPIRE 250, 350 \&  500~\micron data for the DIGIT sources from the Gould Belt Survey \citep[e.g.][]{andre10} and a few other Herschel key programs (KP) and open time (OT) programs \citep{Krause10, stutz10, laun13, tobin10, dun10}. The SPIRE photometry was extracted with DAOPhot within the Herschel Interactive Processing Environment (HIPE) using a top-hat annulus.  In order to derive aperture extraction sizes, we used the Semi-Extended Source Corrector with HIPE.  This routine matches the SPIRE spectral bands by varying the source size.  We then used that source size as our annulus for extraction of SPIRE photometry. The aperture sizes used range from 22$-$47\arcsec. The same apertures were used to extract PACS photometry \citep[for details see][Green et al. in preparation]{green16}. We also made use of the ground based submm/mm photometry for the DIGIT sources presented in \citep{green13}.


\subsubsection{PBRs} \label{sec313}
The photometry of PBRs were obtained from the HOPS catalogue \citep[see][]{stutz13,furlan16}.

%

\subsection{Protostellar properties}
We computed the bolometric luminosity, \lbol, bolometric temperature, \tbol and fractional \smm luminosity, \lbolsmm of the protostars from their SEDs.  \lbol was obtained by integrating under the observed SED over wavelength. We used trapezoidal integration for computing \lbol. The SEDs were extrapolated from the longest observed wavelength (in most cases 850 $\mu$m) as $F_{\nu}~\propto~\nu^2$, before computing \lbol. The bolometric temperature, defined as the temperature of a blackbody with the same mean frequency as the source SED, was computed from the mean frequency of the source SED, following the method of \citet{ml93}. The \smm luminosity, \lsmm was computed by integrating under the SED longward of 350~$\mu$m. The protostellar parameters \lbol, \tbol and \lbolsmm are computed in a uniform way for the HOPS, DIGIT and PBR sources in our sample and are listed in Table~\ref{tab1}. For HOPS and PBR sources which are in Orion, we used a distance of 420~pc  \citep{sandstrom07, menten07, kim08} to compute the luminosities. For the DIGIT sources, the distances listed in  \citet{green13} were used. 

For the HOPS sources, the \lbol~values listed in Table~\ref{tab1} differ from that of \citet{manoj13} by  $\sim$~14\% (median); the median difference in \tbol~is  $\sim$~7\%. These changes can be attributed to the improved photometry used in constructing the observed SED, as described in Section~\ref{sec311}. The \lbol and \tbol values of HOPS and PBR sources listed in Table~1 agree well with those presented in \citet{furlan16}: for most sources they agree within 5-6\%. The major difference is for HOPS 369, which has been modelled as a double source comprising a disk-dominated source and a protostar by \citet{adams12}. We integrated the SED upward of 37~$\micron$  to account only for the protostellar component (see \citet{manoj13}), whereas \citet{furlan16} integrated under the entire SED to obtain \lbol and \tbol. For the DIGIT sources, the  \lbol~ and~\tbol~values listed in Table~\ref{tab1} differ from those in \citet{green13} by $\sim$~15\% (median).  This is because we have used \herschel PACS photometry (70, 100 \& 160~$\micron$) to sample the peak of the SED, while \citet{green13} used PACS spectra, which typically have an absolute flux uncertainty up to $\sim$~30\%, in their SEDs. In summary, we have estimated the protostellar properties for HOPS/PBR and DIGIT sample uniformly, and these quantities agree well with previous estimates within 15-20\%.

As can be seen from Table~\ref{tab1}, a key feature of our sample is the large range (more than 3 order of magnitude) in \lbol, from 0.1 to 275~\lsun. Also, the bolometric temperature, \tbol ranges from 15 to 605~K, indicating that our sample spans a wide range in evolutionary sequence, from early to late protostellar phase.

\begin{deluxetable}{lccccc}  
\tablewidth{0pt}

\tablecaption{Protostellar properties  \label{tab1}}

\tablehead{
\colhead{Object} & \colhead{\lbol} & \colhead{\tbol} & \colhead{\lbolsmm} & \colhead{\lco} & \colhead{\lcolow} \\
\colhead{Name} & \colhead{(\lsun)} & \colhead{(K)} & \colhead{} & \colhead{ ($\times$~\eten{-3} \lsun) } & \colhead{ ($\times$~\eten{-3} \lsun) }
}

\startdata

\cutinhead{HOPS}
HOPS 10  &      3.3 &     46 &     44 &                 1.3 &                0.24\\
HOPS 11  &      8.9 &     47 &     80 &           $\le$~1.2 &          $\le$~0.17\\
HOPS 30  &      3.7 &     77 &     91 &           $\le$~1.5 &          $\le$~0.18\\
HOPS 32  &      2.1 &     60 &    117 &                 1.8 &                0.21\\
HOPS 56  &     22.9 &     45 &    104 &                 8.2 &                1.31\\
HOPS 60  &     22.0 &     55 &    127 &                 6.2 &                0.95\\
HOPS 68  &      5.3 &     88 &     27 &                 1.7 &                0.59\\
HOPS 84  &     49.0 &     90 &    206 &                 0.7 &                0.30\\
HOPS 85  &     16.1 &    166 &    171 &                 1.2 &                0.29\\
HOPS 87  &     36.3 &     38 &     54 &                 7.6 &                0.98\\
HOPS 91  &      4.2 &     37 &     25 &           $\le$~1.3 &          $\le$~0.19\\
HOPS 108 &     36.3 &     34 &     47 &               237.7 &               18.13\\
HOPS 182 &     70.2 &     51 &    157 &                48.4 &                5.08\\
HOPS 203 &     19.8 &     41 &     72 &                 9.7 &                0.82\\
HOPS 288 &    135.6 &     49 &    303 &                10.0 &                1.44\\
HOPS 310 &     13.8 &     51 &    121 &                 9.8 &                1.24\\
HOPS 329 &      2.7 &     79 &     94 &           $\le$~1.3 &          $\le$~0.19\\
HOPS 343 &      3.8 &     79 &    188 &                 2.0 &                0.23\\
HOPS 368 &     63.9 &    150 &    691 &                 6.4 &                1.03\\
HOPS 369 &     18.0 &     35 &     74 &                11.0 &                1.85\\
HOPS 370 &    275.0 &     74 &    697 &                58.0 &                5.24\\

\cutinhead{DIGIT}
IRAS 03245+3002   &      6.6 &     48 &     75 &                 1.3 &                0.20\\
L1455-IRS3        &      0.3 &    236 &     16 &                 0.1 &                0.06\\
IRAS 03301+3111   &      4.5 &    349 &    186 &                 0.9 &                0.09\\
B1-a              &      1.5 &    113 &    285 &                 1.6 &                0.23\\
B1-c              &      3.2 &     46 &    439 &                 0.9 &                0.23\\
L1489             &      3.5 &    248 &    145 &                 0.8 &                0.07\\
IRAM 04191+1522   &      0.1 &     15 &     50 &                 0.2 &                0.03\\
L1551-IRS5        &     22.9 &    108 &    149 &                 1.6 &                0.21\\
L1527             &      1.6 &     79 &     32 &                 0.3 &                0.04\\
TMR 1             &      4.0 &    151 &    191 &                 1.0 &                0.11\\
TMC 1A            &      2.6 &    189 &    120 &                 0.2 &                0.06\\
TMC 1             &      0.7 &    161 &     33 &                 0.6 &                0.05\\
BHR 71            &     11.4 &     45 &     40 &                 9.3 &                0.78\\
DK Cha            &     28.3 &    605 &    655 &                 4.4 &                0.41\\
GSS30-IRS1        &     10.6 &    172 &    432 &                 5.7 &                0.65\\
VLA 1623-243      &      3.3 &     27 &     23 &                 2.0 &                0.31\\
WL12              &      1.6 &    236 &     57 &                 1.1 &                0.10\\
Elias 29          &     15.2 &    310 &    528 &                 6.3 &                0.60\\
B335              &      0.8 &     33 &     20 &                 0.6 &                0.08\\
L1157             &      6.7 &     35 &     30 &                 6.1 &                0.71\\
L1014             &      0.3 &     47 &      9 &           $\le$~0.1 &          $\le$~0.02\\

\cutinhead{PBRs}
HOPS 373 &      5.3 &     37 &     34 &                21.8 &                1.88\\
HOPS 394 &      6.5 &     45 &     38 &                 1.9 &                0.31\\
HOPS 397 &      1.7 &     45 &     40 &                 0.2 &                0.22\\
HOPS 401 &      0.6 &     26 &     13 &           $\le$~1.0 &          $\le$~0.10\\
HOPS 402 &      0.6 &     24 &      9 &           $\le$~1.1 &          $\le$~0.09\\
HOPS 403 &      4.1 &     44 &     34 &                 2.5 &                0.35\\
HOPS 405 &      1.6 &     35 &     28 &                 1.4 &                0.23\\
HOPS 409 &      8.3 &     28 &     47 &                 5.8 &                0.76\\

\enddata
\end{deluxetable}

\begin{figure*}
\includegraphics[scale=0.5]{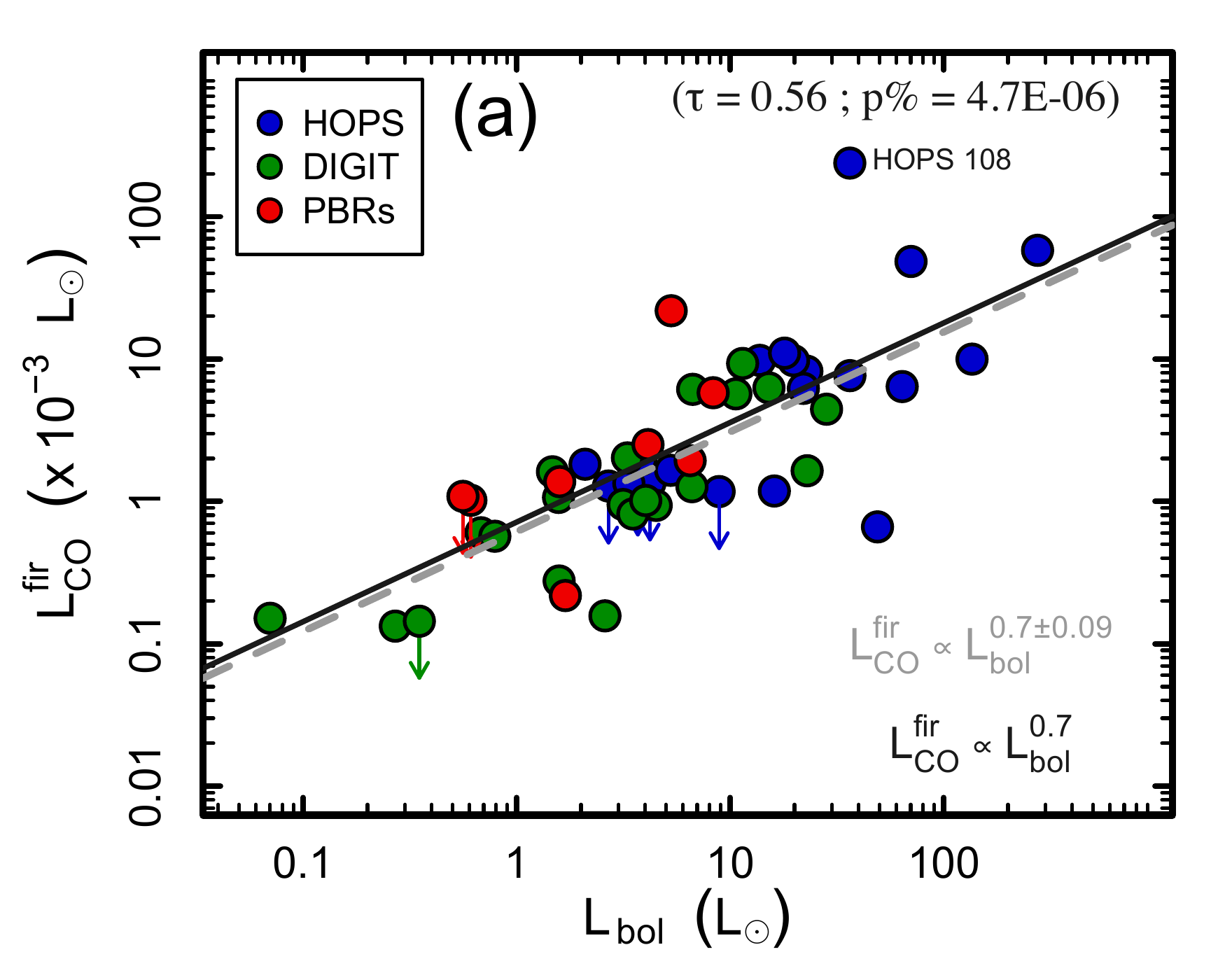}
\includegraphics[scale=0.5]{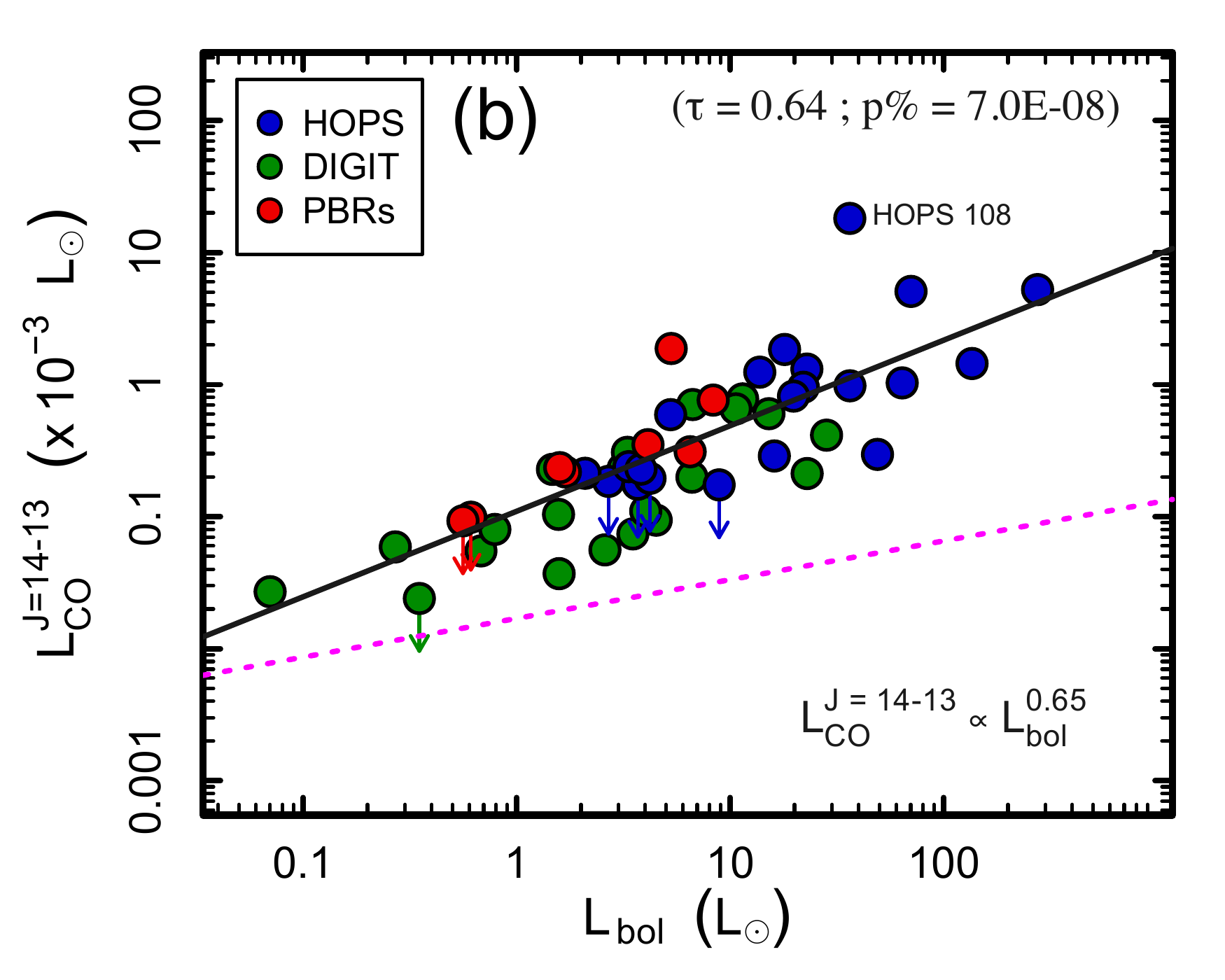}
\caption{ {\bf (a)}~FIR CO luminosity, \lco as a function of \lbol. The HOPS and DIGIT protostars shown as blue and green solid circles respectively, and  the PBRs are shown in red. The downward arrows indicate upper limits in \lco.  The gray dashed line correspond to the best least square fit to the detections only. The black solid line is the Akritas-Thiel-Sen (ATS) line including the censored points.~{\bf (b)}~ Luminosity of the CO (\jj{14}{13}) line, \lcolow  as a function of \lbol. The dashed magenta line indicates the sensitivity of our PACS CO observations converted into luminosity. It has the form \lcolow~$\propto$~\lbol$^{0.3}$ \label{fig1}}
\end{figure*}

\subsection{Far-infrared CO data}
The PACS spectra of the HOPS sources used to measure CO line fluxes were presented in \citet{manoj13}. The CO lines fluxes were measured from the spectra extracted from the central spaxel of the rebinned cube, after applying the PSF loss correction \citep[see][]{manoj13}. The \fir CO luminosity,  \lco, obtained by adding up the luminosities of all the CO lines detected with PACS, and the luminosity of the CO (\jj{14}{13}) line, \lcolow for all the HOPS sources are listed in Table~\ref{tab1}. As before, we used a distance of 420~pc to Orion in computing the luminosities. For the DIGIT sources, the CO line fluxes were measured from the spectra of these sources presented in \citet{green16}.  The \lco and \lcolow for DIGIT sources are shown in Table~\ref{tab1}. The distances to the DIGIT sources were taken from \citet{green13}. The \fir PACS spectra of PBRs were obtained as part of the Herschel open time program OT2\_jtobin\_2. The details of observations and data reductions can be found in Tobin et al (2016). The CO luminosities of the 8 PBRs in our sample are listed in Table~\ref{tab1}.

\section{Results} \label{sec4}

\subsection{Far-IR CO luminosity and \lbol} \label{sec41}

The total \fir~CO luminosity observed with PACS, \lco, is shown as a function of \lbol~in Figure~\ref{fig1}a for HOPS and DIGIT and PBR sources. One of the HOPS sources, HOPS 108, which is in the OMC-2 region, (aka OMC-2 FIR 4) has the highest \lco among our sample. It also has the brightest line spectra of all the protostars in the HOPS sample  \citep{furlan14,manoj13}. However it has been shown recently that the intense line emission seen towards HOPS~108  is not associated with this protostar, but, instead originates in the terminal shock produced by the powerful jet driven by OMC-2 FIR 3 (HOPS 370) \citep{bea16}. Therefore, we do not include HOPS~108 in further analysis. For seven sources, \lco values are upper limits (downward arrows in Figure~\ref{fig1} ): no CO lines are detected in these sources.

Figure~\ref{fig1}a  shows that \lco is strongly correlated with \lbol~over three orders of magnitude in both quantities.  For the detections (non-upper limits), the Pearson's product-moment correlation between log(\lco)~and log(\lbol)~is~0.77 and the associated probability that these two quantities are uncorrelated is $\ll$~\eten{-6}\%. Kendall's rank correlation coefficient, $\tau$, between \lco and \lbol~is~0.56 and the associated probability p\%~$\le$~\eten{-5} and the Spearman's $\rho$~=~0.74 and probability p\%~$\le$~\eten{-6}  indicating that the correlation between \fir~\lco and \lbol~is strong. The functional dependence of~\lco on~\lbol obtained using ordinary least square fit\footnote{We used the {\sc{stats}} package in the R statistical software system \citep{Rpro} to carry out the statistical tests  and least square fit.} (excluding the upper limits in \lco) is found to be \lco~$\propto$~\lbol~$^{0.7\pm0.09}$. 

We also computed the correlation coefficient and the functional dependence of~\lco on~\lbol~including the censored data points (upper limits in \lco). For this we used the Akritas-Thiel-Sen (ATS) regression method, which is the extension of Thiel-Sen regression method \citep{sen68} for censored data \citep{akritas95, fb12}. The Thiel-Sen regression obtains a slope which is the median of the $n(n+1)/2$ slopes of lines defined by all pairs of data points and can be formulated in terms of  Kendall's tau rank correlation coefficient \citep{sen68}. For censored data, pairwise slopes involving censored data points lie in a range of possible values, and the ATS method estimates a distribution function of slopes with these interval-censored values. The median of this distribution becomes a slope estimator for the censored data. The ATS method also provides the generalised Kendall's tau rank correlation coefficient \citep{brown74} for data including censored points \citep[see][]{akritas95, fb12}. 

The generalised Kendall's $\tau$ for the correlation\footnote{Throughout the paper, we have used the {\sc{nada}} package \citep{nada} in R \citep{Rpro} to compute the best fit ATS line and the generalised  Kendall's tau rank correlation coefficient.} between \lco and \lbol is 0.56 and the associated probability is 5.0\ee{-6}\% indicating that the correlation is statistically highly significant. The correlation coefficients and the associated probabilities for the correlations between various quantities are shown in Table~\ref{tab2}. The best fit ATS line which provides the functional dependence of~\lco on~\lbol~including the censored data is shown in Figure~\ref{fig1}a  and has the form 

\[ {\textrm {log \lco}}~=~-3.2 + 0.7\;{\textrm {log \lbol}}\]

\noindent
which can be written as
\[ {\textrm \lco} ~=~{\textrm {6\ee{-4}}} \; {\textrm {\lbol}}^{0.7} \]

\noindent
or equivalently as
\[ {\textrm \lco} / {\textrm \lbol}~=~{\textrm {6\ee{-4}}} \; {\textrm {\lbol}}^{-0.3} \]

\noindent
The \fir~CO luminosity is only a small fraction of the protostellar luminosity \lbol~and the ratio ${\textrm \lco} / {\textrm \lbol}$ drops towards more luminous sources.  ${\textrm \lco} / {\textrm \lbol}$ ranges from 0.4-0.001\% for the protostars in our sample.  Median values of ${\textrm \lco} / {\textrm \lbol}$  as a function of \lbol range from 0.1\% for sources with \lbol < 1~\lsun to 0.02\% for \lbol > 100~\lsun.

\begin{deluxetable*}{lcccccc}[h]  

\tablewidth{0pt}

\tablecaption{Correlation coefficients (Kendall's $\tau$) and associated probabilities \tablenotemark{a}  \label{tab2}}

\tablehead{
\colhead{} & \colhead{\tbol} & \colhead{\lbolsmm} & \colhead{\lco} & \colhead{\lcolow} & \colhead{\lco/\lbol$^{0.7}$} & \colhead{\lcolow/\lbol$^{0.65}$} \\
}

\startdata

\lbol & 0.1~(31\%) & {\bf 0.4~(9\ee{-4}\%)} &{\bf 0.56~(5\ee{-6}\%)} & {\bf 0.64~(7\ee{-8}\%)}& \nodata & \nodata \\
\tbol & \nodata & \nodata &-0.1~(29\%) & \nodata & -0.3~(0.2\%)&-0.3~0.2\%) \\
\lbolsmm & \nodata & \nodata &0.2~(3.1\%) & &  & \\
\lbol$^{0.6}$/\lsmm & \nodata & \nodata  & \nodata & \nodata & -0.2~(11\%)& -0.2~(4\%) \\

\enddata
\tablenotetext{a}{For correlations involving  \lco and \lcolow, which have upper limit points, generalized Kendall's tau and associated probablity are listed. Correlation coefficient and probabilities shown is bold are statistically highly significant.}
\end{deluxetable*}

\begin{figure*}
\includegraphics[scale=0.5]{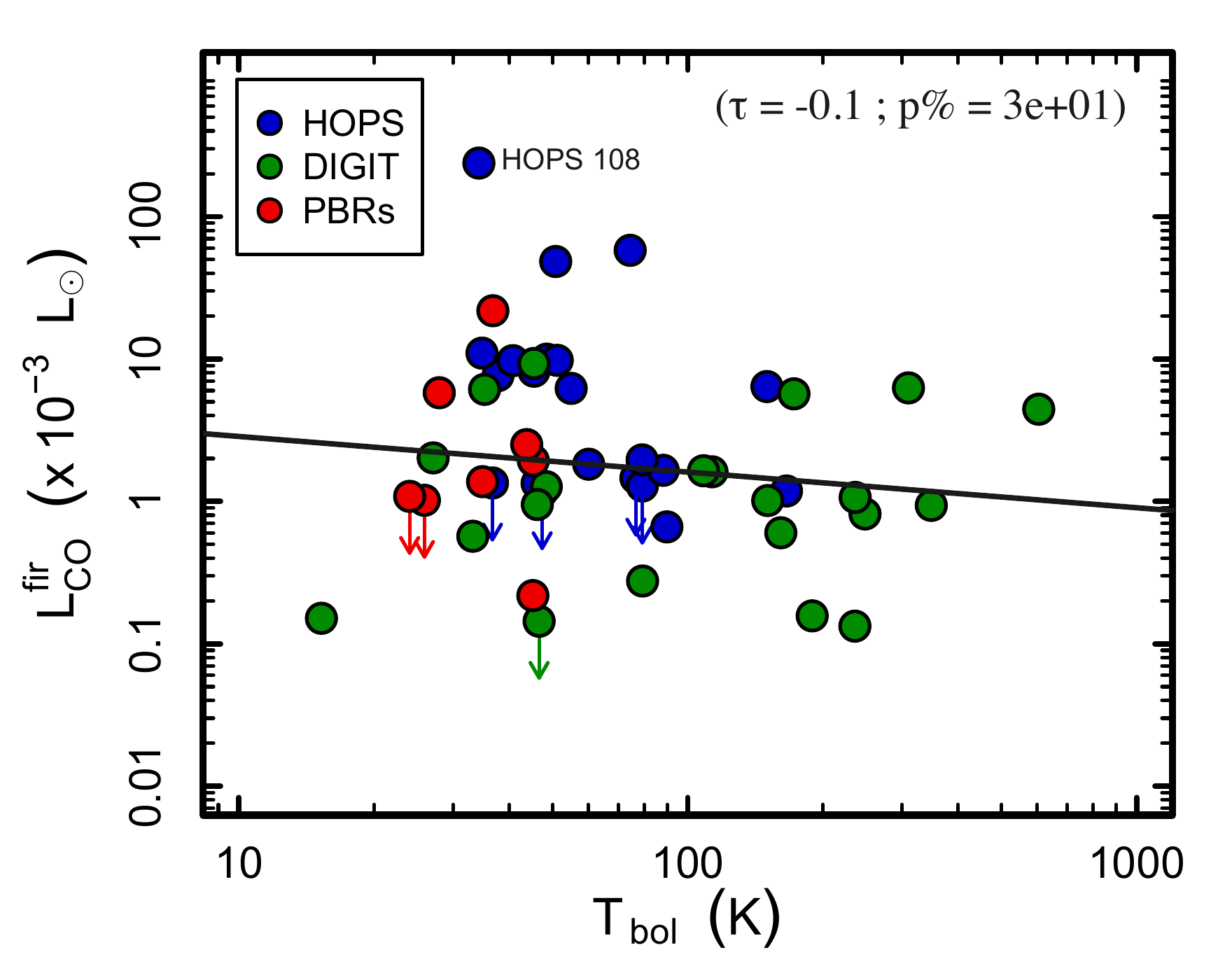}
\includegraphics[scale=0.5]{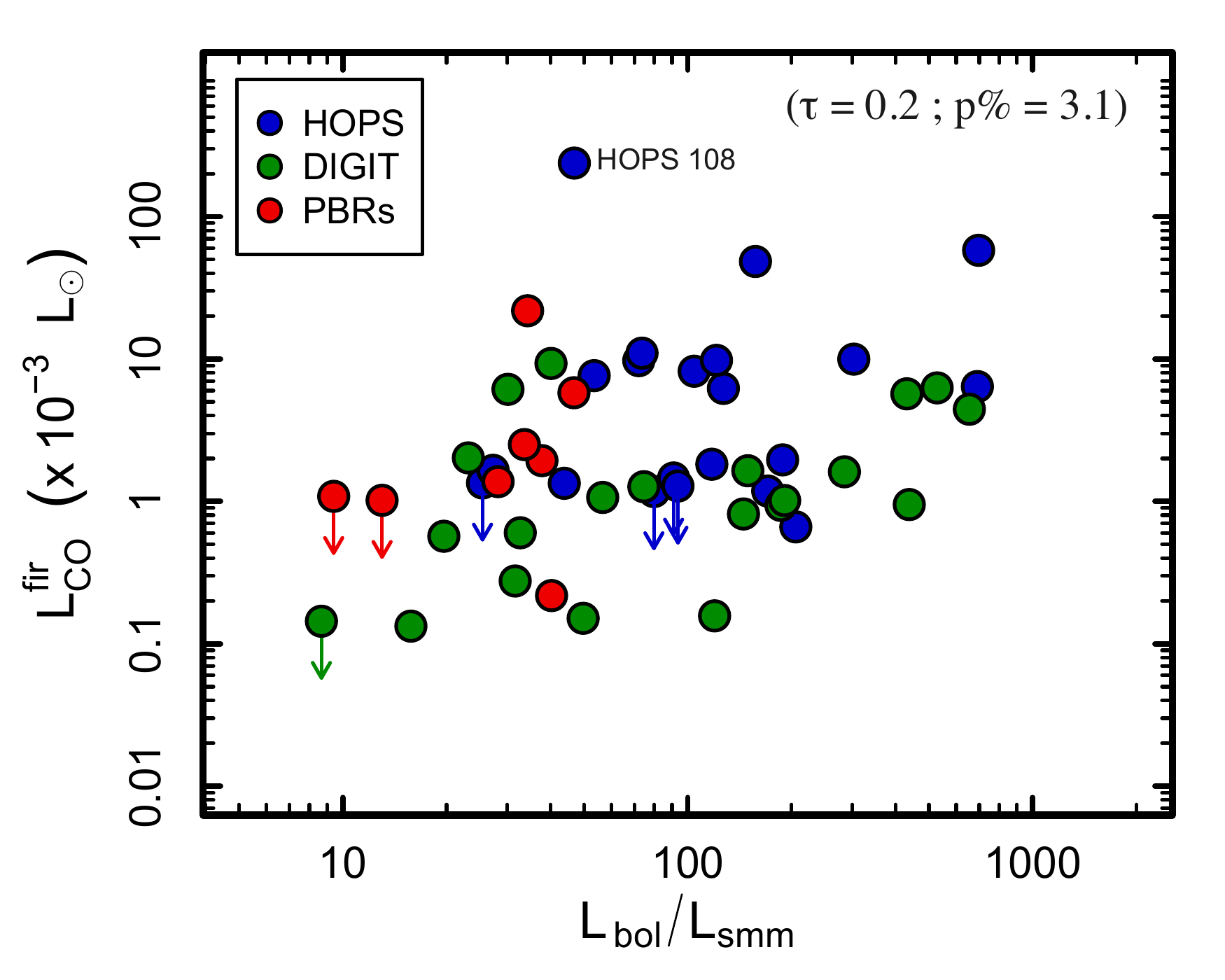}
\caption{FIR CO luminosity as a function of evolutionary indicators \tbol~and \lbolsmm. The HOPS and DIGIT protostars shown as blue and green solid circles respectively, and  the PBRs are shown in red. The downward arrows indicate upper limits in \lco.  \label{fig2}}
\end{figure*}

\subsubsection{Scatter in \lco -- \lbol relation} \label{sec411}

While the correlation between \lco and \lbol is strong and reasonably tight over three orders of magnitude in both the quantities,  there is some spread in the observed \lco values for a given \lbol. The mean absolute deviation from the best fit line in Figure~\ref{fig1} is about a factor of 2 in \lco and the maximum deviation, a factor of 16 lower, is detected for  HOPS~84, which has an \lbol of 49~\lsun but  only three detected CO lines. Part of this scatter is because not all CO lines in the PACS range are detected for all the sources.  \lco is computed by adding up the luminosities of all the detected lines, and, therefore, sources in which more CO lines are detected will have significantly higher \lco compared to those in which very few lines are detected. The scatter produced by this effect can be minimised if luminosity of a single CO line is plotted. To illustrate this,  we show CO (\jj{14}{13}) line luminosity, \lcolow, as a function of \lbol in Figure~\ref{fig1}b. The correlation between \lcolow and \lbol is stronger than that found for \lco: the generalised Kendall's $\tau$ (including the censored data points) for the correlation is 0.64 and the associated probability is 7.0\ee{-8}\%.  The best fit ATS line shown in Figure~\ref{fig1}b has the form  \lcolow~$\propto$~\lbol$^{0.65}$. The correlation between \lcolow and \lbol is also tighter: the mean absolute deviation from the best fit line is only about a factor of 1.6 in \lcolow and the maximum deviation from the fit is only about a factor of 6.  Thus the true dispersion in \lco -- \lbol relation is only a factor of few. To summarise,  \fir CO line luminosities exhibit a strong and tight correlation with \lbol over three orders of magnitude, with an average dispersion of a factor of a few.

\subsubsection{Sensitivity and  the \lco -- \lbol relation} \label{sec412}
The observed 1-$\sigma$ sensitivities of CO line fluxes for the HOPS and DIGIT sample are very similar. For example,  the median CO (\jj{14}{13}) line sensitivity for the HOPS sample is {1.1}\ee{-17} Wm$^{-2}$, while that for the DIGIT sample is {1.3}\ee{-17} Wm$^{-2}$.  However, the HOPS and the PBR sources are in Orion, and are at the same distance (420 pc), while the DIGIT sources are at various distances ranging from 106 to 325 pc. This will result in different sensitivities of CO line luminosities for the HOPS (and PBRs) and the DIGIT sources. The median sensitivity of 
CO (\jj{14}{13}) line luminosity for the DIGIT sources is $\sim$~5 times lower than that for HOPS and PBR sources. In addition, the DIGIT sample preferentially have lower \lbol sources compared to HOPS and PBR sample. The resulting dependence of the sensitivity on \lbol may result in a spurious correlation if there are many non-detections and the detections are just above the sensitivity limit. We show the typical observed sensitivity for the CO (\jj{14}{13}) line as a function of \lbol in Figure~\ref{fig1}b. The observed \lcolow are significantly above the sensitivity for most \lbol. The sensitivity and the sample bias can  at best produce an \lbol dependence of \lcolow~$\propto$~\lbol$^{0.3}$ (the dashed magenta line in Figure~\ref{fig1}b). Thus the correlation between \fir CO luminosities and \lbol is robust.

\begin{figure*}
\includegraphics[scale=0.5]{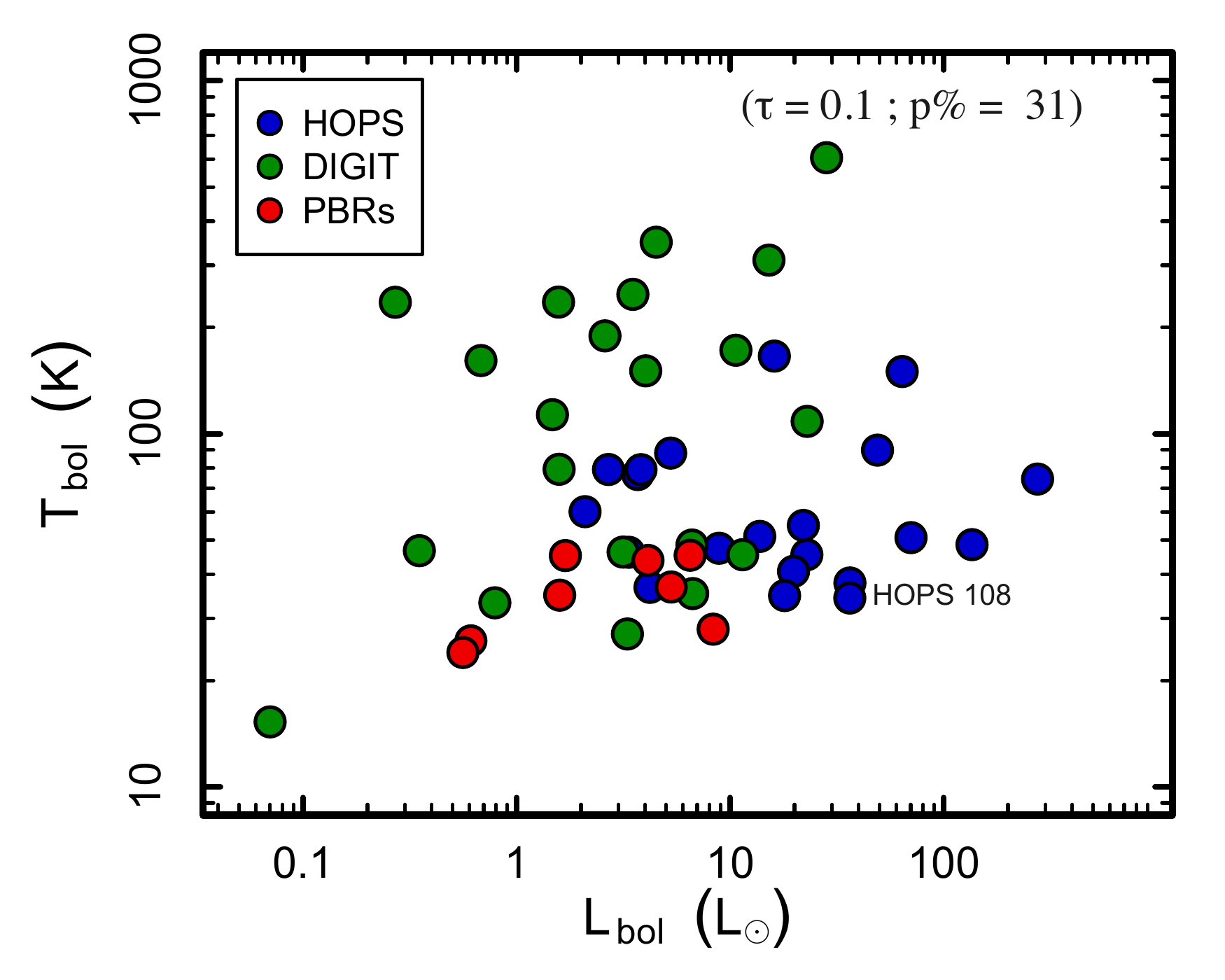}
\includegraphics[scale=0.5]{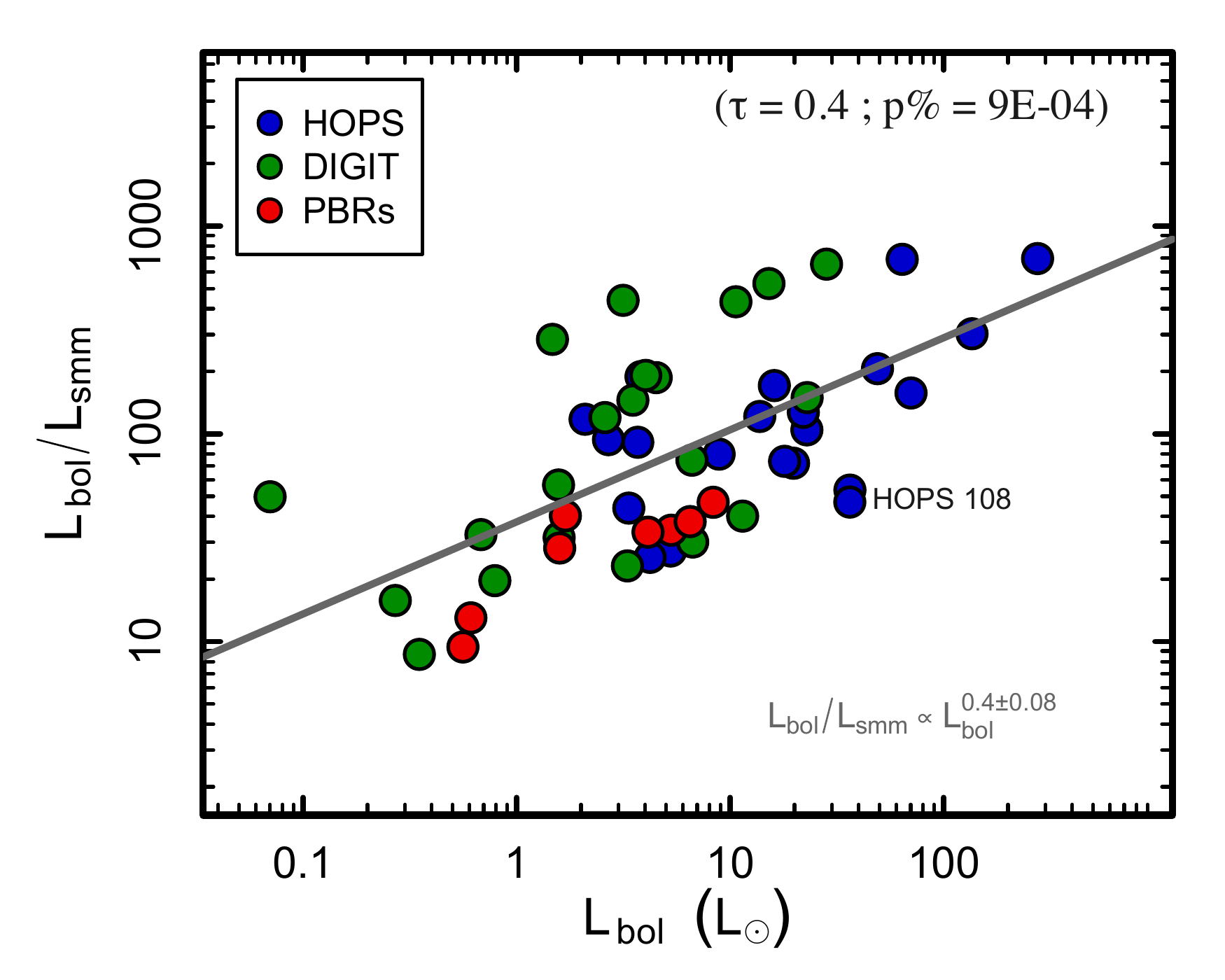}
\caption{ \tbol~and \lbolsmm as functions of \lbol. The HOPS and DIGIT protostars shown as blue and green solid circles respectively, and  the PBRs are shown in red.  \label{fig3}}
\end{figure*}

\subsection{\lco and evolutionary indicators} \label{sec42}
The bolometric temperature, \tbol, and the fractional \smm ~luminosity, \fsmm, are the two commonly used observational tracers of protostellar evolution \citep[e.g.][]{chen95, andre93, andre00, evans09}.  In the following, instead of \fsmm, we use \lbolsmm~so that both \tbol~and \lbolsmm~increases as the protostellar system ages: i.e. more evolved protostars will have larger values of \tbol~and~\lbolsmm.  \lco as a function of \tbol~and~\lbolsmm~is shown in Figure~\ref{fig2}. For the sample of protostars presented here, \fir~\lco does not show any statistically significant correlation with the evolutionary indicators  \tbol~and \lbolsmm.  The generalised Kendall's $\tau$ (including upper limits) for the correlation between \lco and \tbol~is $-$0.1 and the probability that these two quantities are uncorrelated is 29\%.  \lco, on the other hand, show a marginal positive correlation ($\la$~2$\sigma$) with \lbolsmm~($\tau$ = 0.22 \& p = 3.1\%), which seem to suggest 
that \lco increases as the protostar ages, contrary to what is generally expected.  This may result from the strong  \lbol~dependence of \lbolsmm, as can be seen from  Figure~\ref{fig3}. While \tbol~is independent of \lbol~($\tau$=0.1; p=30\%), \lbolsmm~is strongly correlated with \lbol~($\tau$=0.4; p=9\ee{-4}\%) and has a functional dependence of the form \lbol$^{0.4}$.  The  apparent positive  correlation between \lco and \lbolsmm~seen in Figure~\ref{fig2} is caused by the \lbol~dependence of both quantities.

Although both \tbol and \lbolsmm are measures of system age and both are derived from the observed SEDs, \tbol does not show any \lbol dependence while \lbolsmm shows a convincing correlation with \lbol. This is because \lsmm  for our sample sources scales as \lbol$^{0.6}$ and not linearly with \lbol as one would expect. One possible explanation for this is that the outer regions of the protostellar envelopes are also heated externally by the interstellar radiation field, in addition to the heating from the central star. The relative contribution of external heating to \lsmm will be higher in low \lbol sources, thus producing a sub-linear dependence of \lsmm on \lbol. Additionally, the \lbol is highly sensitive to inclination angle \citep[see][]{furlan16}, increasing significantly going from edge-on to face-on viewing geometry. \lsmm, on the other hand, is relatively insensitive to the changes in the viewing angle, resulting in a flatter dependence of \lsmm on \lbol.

\begin{figure*}
\includegraphics[scale=0.5]{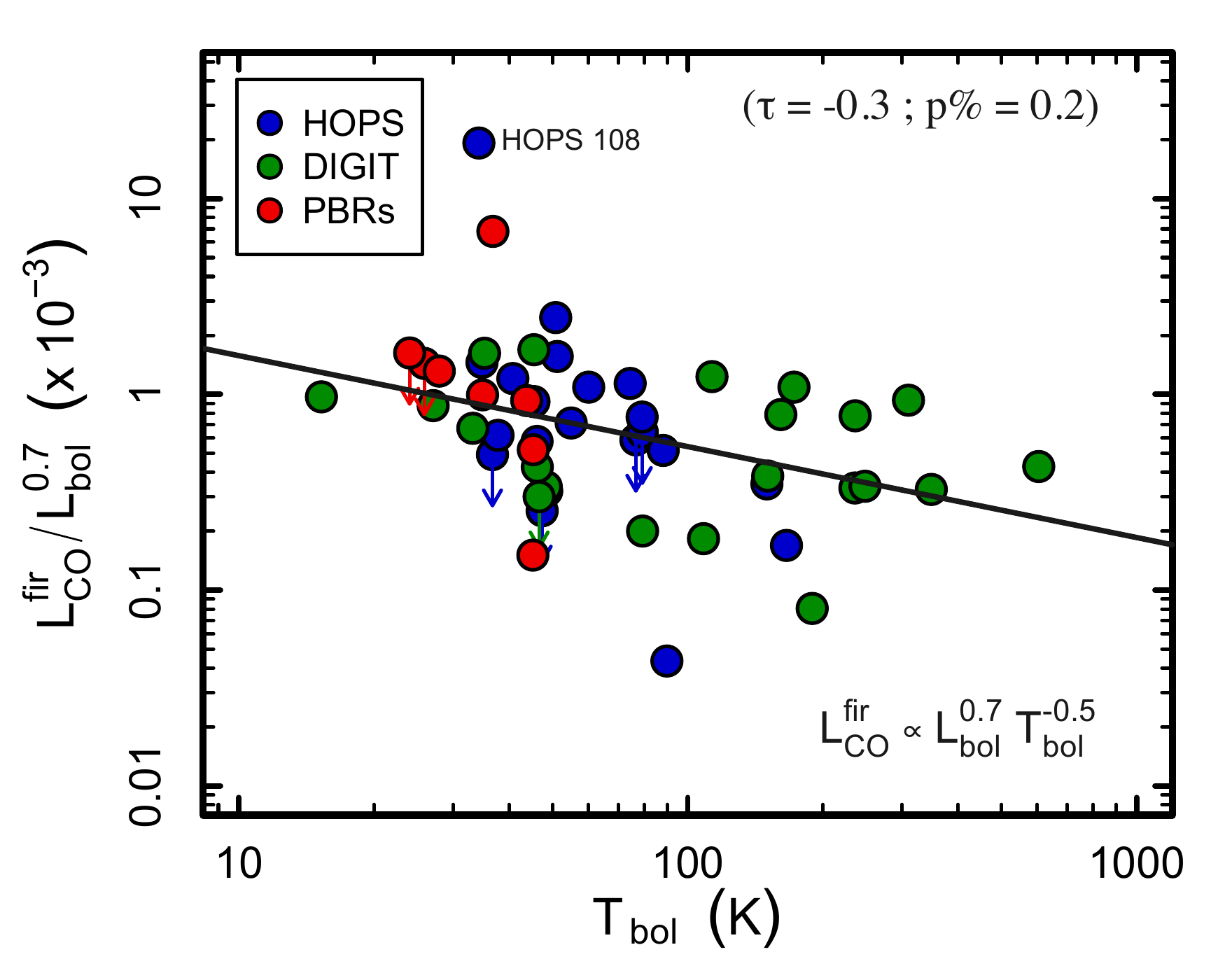}
\includegraphics[scale=0.5]{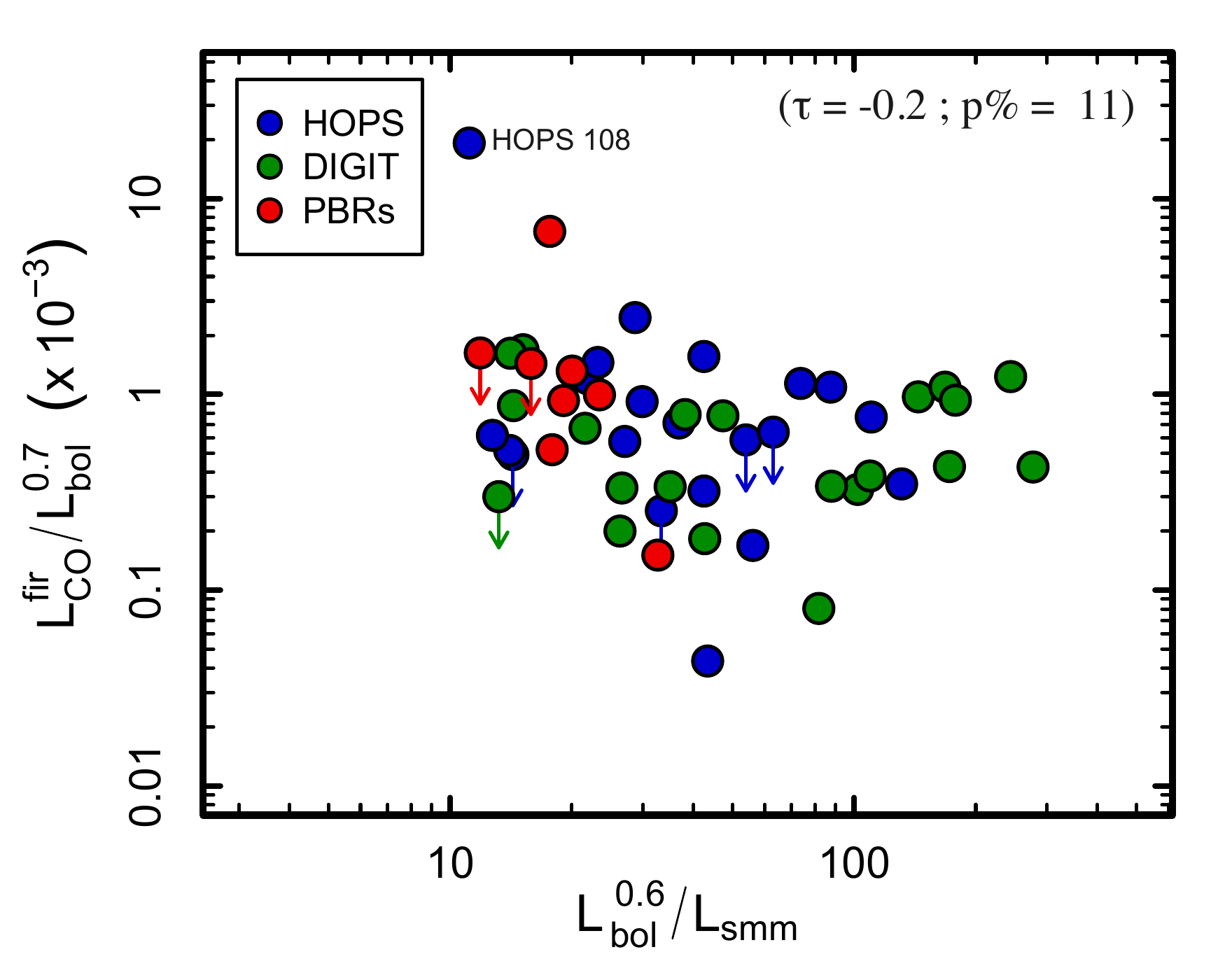}
\caption{ Normalized \lco as a function of \tbol~and \lbolsmm. The HOPS and DIGIT protostars shown as blue and green solid circles respectively, and  the PBRs are shown in red. The downward arrows indicate upper limits in \lco.  \label{fig4}}
\end{figure*}

It is possible that the  strong \lbol~dependence of \lco is masking evolutionary trends in \lco. In order to check this and to emphasize evolutionary trends in \lco, we removed the \lbol~dependence of \lco and \lbolsmm~by normalising them with the functional form of their \lbol~dependence. We then searched for correlations between these normalised quantities as shown in Figure~\ref{fig4}. The normalised \fir CO luminosity, \lco/\lbol$^{0.7}$, which is independent of \lbol, shows a weak  ($\sim$~3$\sigma$) negative correlation with \tbol, indicating that \fir~CO luminosity drops as the protostellar system evolves. The generalised Kendall $\tau$ for the correlation is $-0.3$ and the associated probability is 0.2\%. The ATS non-parametric line fit including the upper limits gives a \tbol~dependence of the form

\[ {\textrm \lco} ~\propto~{\textrm {\lbol}}^{0.7} \; {\textrm {\tbol}}^{-0.5}\]

\noindent
However, \fir~\lco is not correlated with \lbolsmm, even after removing the \lbol~dependence in both the quantities as shown in Figure~\ref{fig4}. The generalised Kendall $\tau$ for the correlation is $-0.2$ and the associated probability is 11\%, indicating that these quantities are uncorrelated.

\begin{figure*}
\includegraphics[scale=0.5]{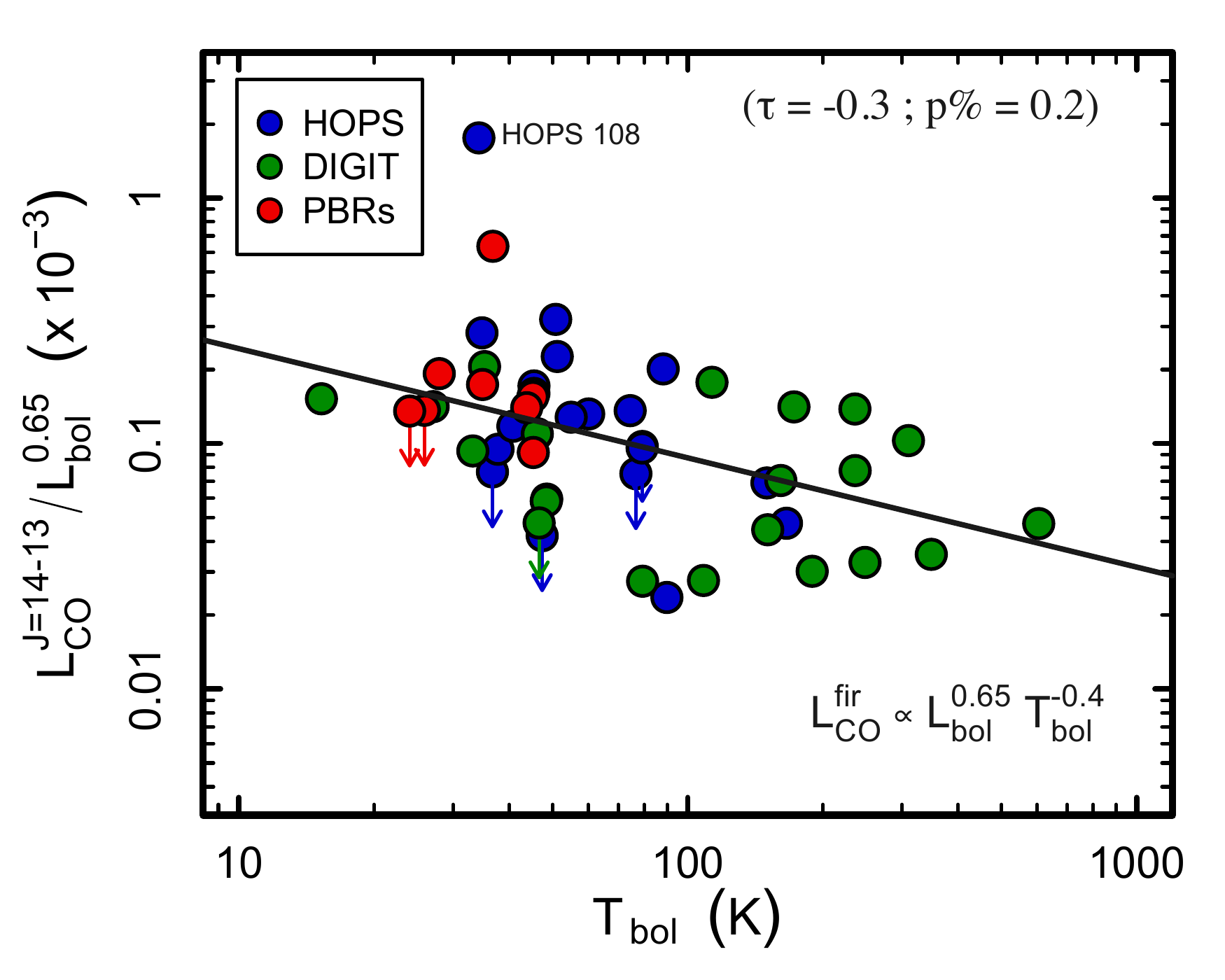}
\includegraphics[scale=0.5]{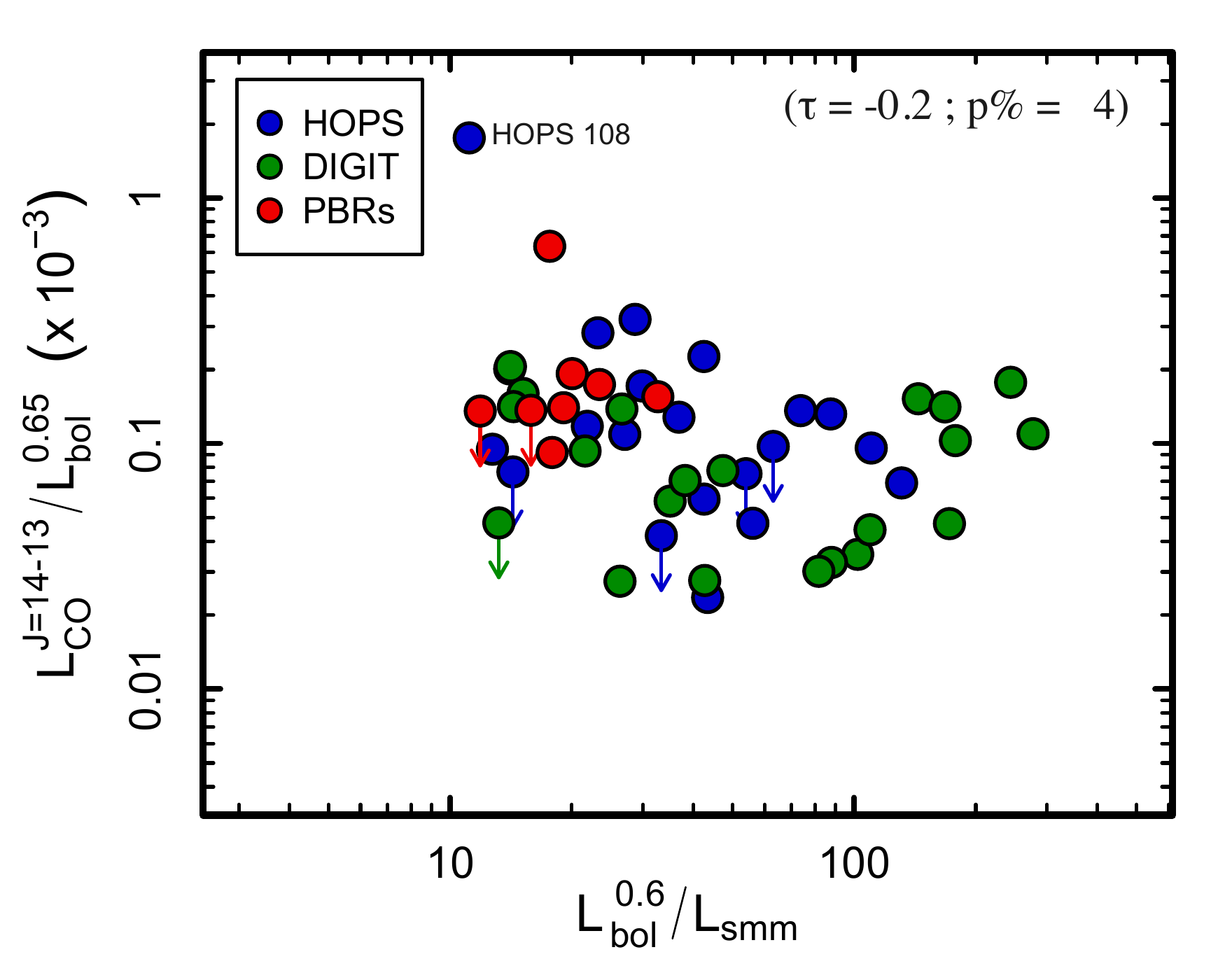}
\caption{ Normalized \lcolow as a function of \tbol~and \lbolsmm. The HOPS and DIGIT protostars shown as blue and green solid circles respectively, and  the PBRs are shown in red. The downward arrows indicate upper limits in \lco.  \label{fig5}}
\end{figure*}

Individual CO line luminosities also display similar behaviour with protostellar evolutionary indicators. In Figure~\ref{fig5} we show \lcolow normalised to its \lbol dependence as a function of \tbol and \lbolsmm normalised to its \lbol dependence. \lcolow/\lbol$^{0.65}$ shows a weak or no correlation with protostellar evolutionary indicators.

\section{Discussion} \label{sec5}

\subsection{Mass accretion-ejection connection in protostars} \label{sec51}
The \fir CO luminosity, \lco is the total luminosity of the high-excitation CO lines in the \fir~(50-200 \micron) observed with PACS onboard~\herschel. \citet{manoj13} have demonstrated that the \fir~CO lines observed towards low-mass protostars arise in hot (T$\ga$2000~K)~gas heated in outflow shocks, a conclusion supported by various other studies with \herschel~\citep[e.g.][]{karska13, matuszak15}. The cooling timescale for the postshock gas at T~$\ga$~2000~K and 
\nhh~$\sim$~\eten{4}$-$\eten{6} \percc is  $<$~100 yr, which is significantly shorter than the protostellar lifetimes \citep[$\sim$~0.5 Myr;][]{dunham14}.  The \fir~CO emission from protostars traces the gas that is currently being shocked by jets/outflows. \lco, then, must be proportional to the total cooling radiation from the shocked gas which, in turn, is proportional to the energy dissipated by jets/outflows. Although other atomic and molecular species such as [O~I], water, and OH can contribute to the cooling in the \fir, \herschel~studies of protostars have shown that \lco is $\sim$~30-40\% of the total \fir-cooling, and, in addition, is proportional to the total \fir-luminosity \citep[Manoj et al. in prep]{karska13, lee14b}.
Thus, \lco can be taken as a lower limit to the mechanical luminosity (\lmech) of the jets/outflows dissipated in shocks over a timescale of $<$~100~yr. In addition, the observed \lco  only includes CO emission from a compact region around the base of the jet/outflow. For Orion sources (HOPS \& PBRs) the spatial extent of the observed CO emission is within a radius of $\sim$~2000~AU  from the protostar \citep[see][]{manoj13}. For DIGIT sources, which are closer, the spatial extent probed is even smaller. The dynamical timescale of such a flow is only 100$-$200~yr for flow velocities of 50$-$~100~\kms. Thus both the cooling timescale and the dynamical timescale of the observed flow are extremely short, and only a small fraction of the protostellar lifetime. \lco, therefore, must be proportional to the instantaneous (smoothed over $\sim$~100~yr) \lmech~=1/2~\mout~${v_{out}}^2$, which ranges over three orders of magnitude for our sample. The jet/outflow velocity, \vout, in protostars are likely to differ by a factor of a few at the most, and the large range in \lmech~(\lco) is primarily due to mass-loss rate, \mout.  

The protostellar luminosity, \lbol~is given by

\[{\textrm \lbol}~=~{\textrm \lphot}~+~{\textrm \lacc}\]

\noindent
where \lphot~is the photospheric luminosity generated by gravitational contraction and deuterium burning and \lacc~is the luminosity released from the accretion of material from disk onto the protostars, which is given by

\[{\textrm \lacc}~=~\eta~\frac{G M(t){\textrm{\macc}}}{r}\]

\noindent
where $\eta$  is the fraction of energy radiated away in the accretion shock, $M(t)$ is the instantaneous protostellar mass and \macc~is the accretion rate onto the protostar. The dominant contribution to \lbol ~is from \lacc~as the contribution from \lphot~becomes important only at late times during protostellar evolution \citep[e.g.][]{tobin12}. The observed points used to construct the SED from which \lbol is computed were measured within a time span of $\sim$~20-30~yr.  Thus \lbol~is a measure of instantaneous (smoothed over $\sim$~30~yr) accretion luminosity, \lacc, which depends on both $M(t)$ and \macc, but the large range (3 orders of magnitude) in \lbol~displayed by our sample sources is primarly driven by \macc.  Therefore, \lbol~should scale with accretion rate, \macc, onto the protostar and the correlation seen in Figure~\ref{fig1} between \lco  and \lbol~is most likely the result of instantaneous \mout~tracking instantaneous \macc. Most jet launching mechanisms proposed for protostars predict such a tight correlation between  \mout and  \macc \citep[e.g.][]{pp92, wk93, shu94}.

The  \lco -- \lbol relation also shows an intrinsic dispersion, as demonstrated in Section~\ref{sec411}.  Various factors could produce such scatter. Although, \lco primarily tracks \mout, it also depends on the flow velocity. \lco could also depend on the properties of the immediate environment around protostars, such as envelope density and ambient gas density. The Orion sources (HOPS \& PBRs) in our sample are from different and diverse regions  within the Orion A (e.g L1641, OMC2/3) and Orion B cloud. The DIGIT sources are from various different star forming regions \citep{green13, green16}. In addition, \lbol has strong inclination angle dependence, while \lco, which represents optically thin CO emission, is less likely to be affected by viewing geometry. All these factors could introduce scatter in the \lco -- \lbol relation. Despite this, the observed dispersion in the \lco -- \lbol relation is quite small: it is only a factor of few compared to the range (three orders of magnitudes) over which the correlation holds. This is possibly because both \lbol and \lco track \macc and \mout smoothed over short and somewhat similar timescales, and, therefore, even if protostellar accretion is episodic, likely measure the same mass accretion/ejection event. Thus, variations in \lbol are closely tracked by \lco, resulting in a strong correlation between the two quantities and a small dispersion.

\subsection{Evolution of mass accretion/ejection in protostars} \label{sec52}
While the observed \lco of protostars is tightly correlated with their luminosity, \lbol, it shows no, or at best a weak, correlation with evolutionary indicators such as \tbol~and \lbolsmm.  As argued in Section~\ref{sec51},  the tight correlation found between \lco and \lbol ~is primarily driven by instantaneous \mout~closely tracking instantaneous \macc. Such a correlation is expected irrespective of the accretion history of protostars and is consistent with both episodic accretion and steady accretion scenarios.  However, we find that \lco, which is proportional to instantaneous \mout,  shows only a marginal correlation with \tbol and no correlation with \lbolsmm. If the accretion is steady, for example, with \macc~monotonically declining with the system age, one would expect a tight correlation between \lco and the evolutionary indicators \tbol~and \lbolsmm. On the other hand, if accretion were episodic, the instantaneous \macc~or \mout~need not have a simple monotonic relation with system age. The lack of strong correlation between \lco and these evolutionary indicators suggests that accretion and jet/outflow activity in protostars are likely episodic.  Moreover, if mass accretion is episodic then \tbol~and \lbolsmm~are not reliable indicators of the physical evolutionary stage of the protostars \citep{dunham10}.  Thus our results are consistent with the mass accretion/ejection in protostars being episodic and do not support a steadily declining \macc~during the protostellar phase.

\subsection{Comparison with (sub)mm CO emission from outflows} \label{sec53}
As pointed out in \S~\ref{sec1}, observations of  low-$J$ ($J_{up}~\leq$ 3) CO lines of the molecular outflows from protostars have shown that the mechanical luminosity, \lcomm, and the momentum flux, \fco~of the molecular flow are correlated with \lbol~\citep[e.g.][]{lada85, cb92, bontemps96, wu04, curtis10}.   \lcomm and \fco are estimated from the total mass in the observed molecular flow and the characteristic velocity of the flow \citep[see][]{cb92, bt99, richer00}.  The kinematical ages (apparent dynamical ages) of the observed molecular flows in low-$J$ ($J_{up}~\leq$ 3) CO lines are in the range of \eten{4}$-$\eten{5}~yr \citep{cb92, bontemps96,bt99}, which is considerably longer than the timescales associated with the protostellar energetics measured by both  \lco and \lbol, and is a significant fraction of the protostellar lifetime. The measured \fco~and \lcomm~are the average momentum and kinetic energy of the molecular flow over the observed dynamical age of the outflow ($\sim$~2$-$20\% of the protostellar lifetime). 

It is interesting to compare the  functional dependence of \lco with \lbol to that observed for \lcomm. The mechanical luminosity of the outflow observed in low-$J$ CO lines, \lcomm  is found to scale with \lbol as \citep{cb92}

\[ {\textrm \lcomm} / {\textrm \lbol}~=~{\textrm {4.3\ee{-2}}} \; {\textrm {\lbol}}^{-0.2} \]

\noindent
Both \lco and \lcomm have very similar dependence on \lbol, over more than three orders of magnitude. However,  \lco is $\sim$~70 times lower than the \lcomm~for a given \lbol~\citep[also see][]{lada85, snell87, wu04}. \lco and \lcomm are measured using independent techniques  at different wavelengths, yet  the remarkable similarity of their \lbol dependence and the large range over which the relation holds suggest that \fir and \smm CO emission are closely connected, and probably caused by a common mechanism.  The molecular outflows are driven by jets from protostars. As the jet propagates, it entrains ambient molecular material and accelerates it, shocking the molecular gas in the process. Part of the kinetic energy of the jet is  dissipated in these shocks and part of it  is transferred to the entrained molecular flow. The \lco traces the currently shocked gas and is a measure of the energy dissipated by the jets in shocks. The \lcomm is the average rate at which kinetic energy is injected into the molecular flow by the jet. Both \lco and \lcomm measure different manifestations of the same process and this possibly explains their similar \lbol dependence.  The observed fact  that \lco is 70 times lower than \lcomm would then mean that more of the kinetic energy in the jet goes into driving the molecular flow than is radiated away in shocks.


The measured \fco~and \lcomm~are 
proportional to the mass ejection rate from the protostar smoothed over \eten{4}$-$\eten{5}~yr, or, in other words, the time-averaged mass ejection rate from the protostar, \moutav.  
The average momentum flux of the molecular outflow \fco~is found to be correlated with protostellar evolutionary indicators. \citet{bontemps96} have found that \fco~is roughly proportional to the envelope mass, \menv, which they interpret as a progressive decline of outflow activity (or equivalently mass ejection/accretion rate) during the protostellar accretion phase.  \citet{curtis10} have confirmed the decline of \fco~with decreasing \menv~and, in addition, find that \fco~drops with increasing \tbol~with a functional for \fco~$\sim$~\tbol$^{-0.6}$. These results indicate that both~\moutav~and~\maccav~steadily drop with protostellar age. Our results, however, show that \lco is only weakly correlated with \tbol and not correlated with \lbolsmm, indicating that the instantaneous mass loss rate, \mout, as measured by \fir~\lco does not show clear evidence of a steady decline with protostellar evolutionary tracers. This suggests that the instantaneous \mout~and \macc~are not  monotonically decreasing functions of system age. One possible explanation for this behaviour is that the instantaneous accretion/ejection rate onto protostars is highly time variable and episodic, but the amplitude and/or frequency of this variability decreases with time such that the time averaged accretion/ejection rates decline with protostellar age. Numerical simulations of the accretion history of protostars are consistent with such a time evolution for protostellar accretion rates \citep[e.g][]{vb08, vb10, machida13}. 

We note, however, that the studies of the energetics derived from the low-$J$ CO lines in the \submm and the high-excitation CO lines in the \fir are carried out on different samples. These samples differ in the distribution of protostellar properties, and in some cases, the observed SEDs from which the protostellar properties are estimated are not well sampled. For example, our sample has \lbol ranging from 0.07 to 275~\lsun, with a median value of $\sim$~5~\lsun, whereas the sample of \citet{bontemps96} has much narrower range in \lbol (0.2$-$41~\lsun), with a slightly lower median of 2~\lsun.  Although the \citet{curtis10} sample has an \lbol range (0.03$-$100~\lsun) and median similar to our sample, the observed SEDs from which protostellar properties are computed are sparsely sampled, particularly the peak of the SED \citep[see][]{hatchell07b}. Thus the correlations discussed above between various quantities will have to be demonstrated for the same sample, whose protostellar properties are well characterised, before one can draw robust conclusions about the time evolution of mass accretion/ejection in protostars.

\section{Summary \& Conclusions}
We studied the evolution of \fir, high$-J$ ($14~\leq~J_{up}~\leq~45$)  CO emission from protostars by analysing the \herschel/PACS spectra of 50 embedded sources in the nearby star forming regions observed by the HOPS and DIGIT \herschel key programmes.  We first constructed uniformly sampled SEDs for these sources. The peak of the SEDs are well sampled with \herschel/PACS photometric observations at 70, 100 \& 160~$\micron$. We then computed the bolometric luminosity, \lbol, bolometric temperature, \tbol and fractional \smm luminosity, \lbolsmm of the protostars in our sample in a uniform way from the observed SEDs. The protostars in our sample have a large range in \lbol (more than 3 orders of magnitude)  and are at various stages of evolution. For this sample, we searched for correlations between \fir CO line luminosities and various protostellar properties. Our main results and conclusions are summarised below.

\begin{itemize}
\item 
 We find a strong and tight correlation between  \fir CO luminosity, \lco and the bolometric luminosity, \lbol of the protostars. The \fir CO luminosity, \lco scales with \lbol as \lco~$\propto$~\lbol$^{0.7}$. This correlation extends over more than three orders of magnitude in both quantities, with a mean dispersion from the relation of  less than a factor of 2.

\item
We find a weak correlation between \lco and \tbol, but no correlation is found between \lco and  \lbolsmm. 

\item
FIR~CO emission from protostars trace the currently shocked gas by jets/outflows, the cooling timescales for which are $<$~100 yr, significantly shorter than the protostellar lifetimes. \lco, is proportional to the instantaneous mechanical luminosity of the jet/outflow, which scales with instantaneous mass loss rate, \mout. The correlation between \lco and \lbol, then, is  indicative of instantaneous \mout  tracking instantaneous \macc.

\item
The lack of (or weak) correlation between \lco and evolutionary indicators \tbol and \lbolsmm suggests that \mout and, therefore,  \macc do not show any clear evolutionary trend. Thus our results are consistent with mass accretion/ejection in protostars being episodic. 

\item
We compared our results with those found for the the mechanical luminosity (\lmech) and the momentum flux or outflow force (\fco) of the molecular outflows observed in low-$J$ ($J_{up}~\leq$ 3) CO lines at (sub-)mm wavelengths. The functional dependence of \lco on \lbol that we find is similar to that found for the mechanical luminosity, \lcomm, of molecular outflows observed in low-excitation CO lines. The observed similarity  and the large range over which the relations hold suggest that \fir and (sub)mm CO emission are closely connected. 

Studies of  molecular outflows in low-$J$ ($J_{up}~\leq$ 3) also indicate that the time-averaged mass ejection/accretion rate steadily declines during the protostellar phase \citep{bontemps96}. Our results, on the other hand, suggests that the instantaneous accretion/ejection rate does not show clear evolutionary trend. One possible explanation for this is that mass accretion/ejection  rate in protostars is highly time variable and episodic, but the amplitude and/or frequency of this variability decreases with time such that the time averaged accretion/ejection rate declines with system age.

These correlations will have to be demonstrated for the same sample from a homogeneous set of observations before the detailed behaviour of the time evolution of mass accretion/ejection in protostars can be confirmed.

\end{itemize}

\acknowledgments
Support for this work, part of the Herschel Open Time Key Project Program, was provided by NASA through an award issued by the Jet Propulsion Laboratory, California Institute of Technology. This work was supported by NSF grant AST-1109116 to the University of Texas at Austin. This work is based on observations made with the {\it Herschel Space Observatory}, a European Space
Agency Cornerstone Mission with significant participation by NASA; it is also on observations made with the Spitzer Space Telescope, which is operated by the Jet Propulsion Laboratory (JPL), California Institute of Technology (Caltech), under a contract with NASA. We also include data from the Atacama Pathfinder Experiment, a collaboration between the Max-Planck Institut f\"{u}r Radio-astronomie, the European Southern Observatory, and the Onsala Space Observatory. This publication makes use of data products from the Two Micron All Sky Survey, which is a joint project of the University of Massachusetts and the Infrared Processing and Analysis Center/Caltech, funded by NASA and the NSF.


\begin{thebibliography}{}
\expandafter\ifx\csname natexlab\endcsname\relax\def\natexlab#1{#1}\fi

\bibitem[{{Adams} {et~al.}(2012){Adams}, {Herter}, {Osorio}, {Macias},
  {Megeath}, {Fischer}, {Ali}, {Calvet}, {D'Alessio}, {De Buizer}, {Gull},
  {Henderson}, {Keller}, {Morris}, {Remming}, {Schoenwald}, {Shuping},
  {Stacey}, {Stanke}, {Stutz}, \& {Vacca}}]{adams12}
{Adams}, J.~D., {Herter}, T.~L., {Osorio}, M., {et~al.} 2012, \apjl, 749, L24

\bibitem[{{Akritas} {et~al.}(1995){Akritas}, {Murphy}, \&
  {Lavalley}}]{akritas95}
{Akritas}, M.~G., {Murphy}, S.~A., \& {Lavalley}, M.~P. 1995, Journal of the
  American Statistical Association, 90, 170

\bibitem[{{Andr\'{e}} {et~al.}(1993){Andr\'{e}}, {Ward-Thompson}, \&
  {Barsony}}]{andre93}
{Andr\'{e}}, P., {Ward-Thompson}, D., \& {Barsony}, M. 1993, \apj, 406, 122

\bibitem[{{Andr\'{e}} {et~al.}(2000){Andr\'{e}}, {Ward-Thompson}, \&
  {Barsony}}]{andre00}
---. 2000, Protostars and Planets IV, 59

\bibitem[{{Andr{\'e}} {et~al.}(2010){Andr{\'e}}, {Men'shchikov}, {Bontemps},
  {K{\"o}nyves}, {Motte}, {Schneider}, {Didelon}, {Minier}, {Saraceno},
  {Ward-Thompson}, {di Francesco}, {White}, {Molinari}, {Testi}, {Abergel},
  {Griffin}, {Henning}, {Royer}, {Mer{\'{\i}}n}, {Vavrek}, {Attard},
  {Arzoumanian}, {Wilson}, {Ade}, {Aussel}, {Baluteau}, {Benedettini},
  {Bernard}, {Blommaert}, {Cambr{\'e}sy}, {Cox}, {di Giorgio}, {Hargrave},
  {Hennemann}, {Huang}, {Kirk}, {Krause}, {Launhardt}, {Leeks}, {Le Pennec},
  {Li}, {Martin}, {Maury}, {Olofsson}, {Omont}, {Peretto}, {Pezzuto}, {Prusti},
  {Roussel}, {Russeil}, {Sauvage}, {Sibthorpe}, {Sicilia-Aguilar}, {Spinoglio},
  {Waelkens}, {Woodcraft}, \& {Zavagno}}]{andre10}
{Andr{\'e}}, P., {Men'shchikov}, A., {Bontemps}, S., {et~al.} 2010, \aap, 518,
  L102

\bibitem[{{Arce} {et~al.}(2007){Arce}, {Shepherd}, {Gueth}, {Lee}, {Bachiller},
  {Rosen}, \& {Beuther}}]{arce07}
{Arce}, H.~G., {Shepherd}, D., {Gueth}, F., {et~al.} 2007, Protostars and
  Planets V, 245

\bibitem[{{Audard} {et~al.}(2014){Audard}, {{\'A}brah{\'a}m}, {Dunham},
  {Green}, {Grosso}, {Hamaguchi}, {Kastner}, {K{\'o}sp{\'a}l}, {Lodato},
  {Romanova}, {Skinner}, {Vorobyov}, \& {Zhu}}]{audard14}
{Audard}, M., {{\'A}brah{\'a}m}, P., {Dunham}, M.~M., {et~al.} 2014, Protostars
  and Planets VI, 387

\bibitem[{{Bachiller} \& {Tafalla}(1999)}]{bt99}
{Bachiller}, R., \& {Tafalla}, M. 1999, in NATO Advanced Science Institutes
  (ASI) Series C, Vol. 540, NATO Advanced Science Institutes (ASI) Series C,
  ed. C.~J. {Lada} \& N.~D. {Kylafis}, 227

\bibitem[{{Bally} \& {Lada}(1983)}]{bl83}
{Bally}, J., \& {Lada}, C.~J. 1983, \apj, 265, 824

\bibitem[{{Bally} {et~al.}(2007){Bally}, {Reipurth}, \& {Davis}}]{bally07}
{Bally}, J., {Reipurth}, B., \& {Davis}, C.~J. 2007, Protostars and Planets V,
  215

\bibitem[{{Bontemps} {et~al.}(1996){Bontemps}, {Andr\'{e}}, {Terebey}, \&
  {Cabrit}}]{bontemps96}
{Bontemps}, S., {Andr\'{e}}, P., {Terebey}, S., \& {Cabrit}, S. 1996, \aap,
  311, 858

\bibitem[{{Brown} {et~al.}(1974){Brown}, {Hollander}, \& {Korwar}}]{brown74}
{Brown}, B.~W., {Hollander}, M., \& {Korwar}, R.~M. 1974, in Reliability and
  Biometry, ed. F.~{Proschan} \& R.~J. {Serfling} (SIAM, Philadelphia),
  327--354

\bibitem[{{Cabrit} \& {Bertout}(1992)}]{cb92}
{Cabrit}, S., \& {Bertout}, C. 1992, \aap, 261, 274

\bibitem[{{Calvet} \& {Gullbring}(1998)}]{cg98}
{Calvet}, N., \& {Gullbring}, E. 1998, \apj, 509, 802

\bibitem[{{Calvet} {et~al.}(2004){Calvet}, {Muzerolle}, {Brice{\~n}o},
  {Hern{\'a}ndez}, {Hartmann}, {Saucedo}, \& {Gordon}}]{calvet04}
{Calvet}, N., {Muzerolle}, J., {Brice{\~n}o}, C., {et~al.} 2004, \aj, 128, 1294

\bibitem[{{Chen} {et~al.}(1995){Chen}, {Myers}, {Ladd}, \& {Wood}}]{chen95}
{Chen}, H., {Myers}, P.~C., {Ladd}, E.~F., \& {Wood}, D.~O.~S. 1995, \apj, 445,
  377

\bibitem[{{Cieza} {et~al.}(2013){Cieza}, {Olofsson}, {Harvey}, {Evans},
  {Najita}, {Henning}, {Mer{\'{\i}}n}, {Liebhart}, {G{\"u}del}, {Augereau}, \&
  {Pinte}}]{cieza13}
{Cieza}, L.~A., {Olofsson}, J., {Harvey}, P.~M., {et~al.} 2013, \apj, 762, 100

\bibitem[{{Curtis} {et~al.}(2010){Curtis}, {Richer}, {Swift}, \&
  {Williams}}]{curtis10}
{Curtis}, E.~I., {Richer}, J.~S., {Swift}, J.~J., \& {Williams}, J.~P. 2010,
  \mnras, 408, 1516

\bibitem[{{Dionatos} {et~al.}(2013){Dionatos}, {J{\o}rgensen}, {Green},
  {Herczeg}, {Evans}, {Kristensen}, {Lindberg}, \& {van Dishoeck}}]{dionatos13}
{Dionatos}, O., {J{\o}rgensen}, J.~K., {Green}, J.~D., {et~al.} 2013, \aap,
  558, A88

\bibitem[{{Dunham}(2010)}]{dun10}
{Dunham}, M. 2010, {OT1\_mdunham\_1: Understanding the Protostellar Mass
  Accretion Process: Herschel 100-500 micron Photometry of Low Luminosity
  Embedded Protostars}, Herschel Space Observatory Proposal, id.1359, ,

\bibitem[{{Dunham} {et~al.}(2014{\natexlab{a}}){Dunham}, {Arce}, {Mardones},
  {Lee}, {Matthews}, {Stutz}, \& {Williams}}]{dunham14b}
{Dunham}, M.~M., {Arce}, H.~G., {Mardones}, D., {et~al.} 2014{\natexlab{a}},
  \apj, 783, 29

\bibitem[{{Dunham} {et~al.}(2010){Dunham}, {Evans}, {Terebey}, {Dullemond}, \&
  {Young}}]{dunham10}
{Dunham}, M.~M., {Evans}, II, N.~J., {Terebey}, S., {Dullemond}, C.~P., \&
  {Young}, C.~H. 2010, \apj, 710, 470

\bibitem[{{Dunham} \& {Vorobyov}(2012)}]{dunham12}
{Dunham}, M.~M., \& {Vorobyov}, E.~I. 2012, \apj, 747, 52

\bibitem[{{Dunham} {et~al.}(2014{\natexlab{b}}){Dunham}, {Stutz}, {Allen},
  {Evans}, {Fischer}, {Megeath}, {Myers}, {Offner}, {Poteet}, {Tobin}, \&
  {Vorobyov}}]{dunham14}
{Dunham}, M.~M., {Stutz}, A.~M., {Allen}, L.~E., {et~al.} 2014{\natexlab{b}},
  Protostars and Planets VI, 195

\bibitem[{{Evans} {et~al.}(2009){Evans}, {Dunham}, {J{\o}rgensen}, {Enoch},
  {Mer{\'{\i}}n}, {van Dishoeck}, {Alcal{\'a}}, {Myers}, {Stapelfeldt},
  {Huard}, {Allen}, {Harvey}, {van Kempen}, {Blake}, {Koerner}, {Mundy},
  {Padgett}, \& {Sargent}}]{evans09}
{Evans}, N.~J., {Dunham}, M.~M., {J{\o}rgensen}, J.~K., {et~al.} 2009, \apjs,
  181, 321

\bibitem[{{Fedele} {et~al.}(2013){Fedele}, {Bruderer}, {van Dishoeck}, {Carr},
  {Herczeg}, {Salyk}, {Evans}, {Bouwman}, {Meeus}, {Henning}, {Green},
  {Najita}, \& {G{\"u}del}}]{Fedele13}
{Fedele}, D., {Bruderer}, S., {van Dishoeck}, E.~F., {et~al.} 2013, \aap, 559,
  A77

\bibitem[{{Feigelson} \& {Babu}(2012)}]{fb12}
{Feigelson}, E.~D., \& {Babu}, G.~J. 2012, {Modern Statistical Methods for
  Astronomy} (Cambridge University Press)

\bibitem[{{Fischer} {et~al.}(2010){Fischer}, {Megeath}, {Ali}, {Tobin},
  {Osorio}, {Allen}, {Kryukova}, {Stanke}, {Stutz}, {Bergin}, {Calvet}, {di
  Francesco}, {Furlan}, {Hartmann}, {Henning}, {Krause}, {Manoj}, {Maret},
  {Muzerolle}, {Myers}, {Neufeld}, {Pontoppidan}, {Poteet}, {Watson}, \&
  {Wilson}}]{fischer10}
{Fischer}, W.~J., {Megeath}, S.~T., {Ali}, B., {et~al.} 2010, \aap, 518, L122

\bibitem[{{Fischer} {et~al.}(2012){Fischer}, {Megeath}, {Tobin}, {Stutz},
  {Ali}, {Remming}, {Kounkel}, {Stanke}, {Osorio}, {Henning}, {Manoj}, \&
  {Wilson}}]{fischer12}
{Fischer}, W.~J., {Megeath}, S.~T., {Tobin}, J.~J., {et~al.} 2012, \apj, 756,
  99

\bibitem[{{Fischer} {et~al.}(2013){Fischer}, {Megeath}, {Stutz}, {Tobin},
  {Ali}, {Stanke}, {Osorio}, {Furlan}, {HOPS Team}, \& {Orion Protostar
  Survey}}]{fischer13}
{Fischer}, W.~J., {Megeath}, S.~T., {Stutz}, A.~M., {et~al.} 2013,
  Astronomische Nachrichten, 334, 53

\bibitem[{{Frank} {et~al.}(2014){Frank}, {Ray}, {Cabrit}, {Hartigan}, {Arce},
  {Bacciotti}, {Bally}, {Benisty}, {Eisl{\"o}ffel}, {G{\"u}del}, {Lebedev},
  {Nisini}, \& {Raga}}]{frank14}
{Frank}, A., {Ray}, T.~P., {Cabrit}, S., {et~al.} 2014, Protostars and Planets
  VI, 451

\bibitem[{{Furlan} {et~al.}(2014){Furlan}, {Megeath}, {Osorio}, {Stutz},
  {Fischer}, {Ali}, {Stanke}, {Manoj}, {Adams}, \& {Tobin}}]{furlan14}
{Furlan}, E., {Megeath}, S.~T., {Osorio}, M., {et~al.} 2014, \apj, 786, 26

\bibitem[{{Furlan} {et~al.}(2016){Furlan}, {Fischer}, {Ali}, {Stutz}, {Stanke},
  {Tobin}, {Megeath}, {Osorio}, {Hartmann}, {Calvet}, {Poteet}, {Booker},
  {Manoj}, {Watson}, \& {Allen}}]{furlan16}
{Furlan}, E., {Fischer}, W.~J., {Ali}, B., {et~al.} 2016, \apjs, 224, 5

\bibitem[{{Gonzalez-Garcia} {et~al.}(2016){Gonzalez-Garcia}, {Manoj}, {Watson},
  {Vavrek}, {Megeath}, {Stutz}, {Osorio}, {Wyrowski}, {Fischer}, {Tobin},
  {Sanchez-Portal}, {Diaz Rodriguez}, \& {Wilson}}]{bea16}
{Gonzalez-Garcia}, B., {Manoj}, P., {Watson}, D.~M., {et~al.} 2016, \aap,
  Submitted to A \& A

\bibitem[{{Green} {et~al.}(2006){Green}, {Hartmann}, {Calvet}, {Watson},
  {Ibrahimov}, {Furlan}, {Sargent}, \& {Forrest}}]{green06}
{Green}, J.~D., {Hartmann}, L., {Calvet}, N., {et~al.} 2006, \apj, 648, 1099

\bibitem[{{Green} {et~al.}(2013{\natexlab{a}}){Green}, {Evans},
  {K{\'o}sp{\'a}l}, {Herczeg}, {Quanz}, {Henning}, {van Kempen}, {Lee},
  {Dunham}, {Meeus}, {Bouwman}, {Chen}, {G{\"u}del}, {Skinner}, {Liebhart}, \&
  {Merello}}]{green13b}
{Green}, J.~D., {Evans}, II, N.~J., {K{\'o}sp{\'a}l}, {\'A}., {et~al.}
  2013{\natexlab{a}}, \apj, 772, 117

\bibitem[{{Green} {et~al.}(2013{\natexlab{b}}){Green}, {Evans}, {J{\o}rgensen},
  {Herczeg}, {Kristensen}, {Lee}, {Dionatos}, {Yildiz}, {Salyk}, {Meeus},
  {Bouwman}, {Visser}, {Bergin}, {van Dishoeck}, {Rascati}, {Karska}, {van
  Kempen}, {Dunham}, {Lindberg}, {Fedele}, \& {DIGIT Team}}]{green13}
{Green}, J.~D., {Evans}, II, N.~J., {J{\o}rgensen}, J.~K., {et~al.}
  2013{\natexlab{b}}, \apj, 770, 123

\bibitem[{{Green} {et~al.}(2016){Green}, {Yang}, {Evans}, {Karska}, {Herczeg},
  {van Dishoeck}, {Lee}, {Larson}, \& {Bouwman}}]{green16}
{Green}, J.~D., {Yang}, Y.-L., {Evans}, II, N.~J., {et~al.} 2016, ArXiv
  e-prints, arXiv:1601.05028

\bibitem[{{Gullbring} {et~al.}(1998){Gullbring}, {Hartmann}, {Briceno}, \&
  {Calvet}}]{ghbc98}
{Gullbring}, E., {Hartmann}, L., {Briceno}, C., \& {Calvet}, N. 1998, \apj,
  492, 323

\bibitem[{{Hartmann}(1998)}]{hart98}
{Hartmann}, L. 1998, {Accretion processes in star formation} (Cambridge, UK ;
  New York : Cambridge University Press, 1998.~(Cambridge astrophysics series ;
  32))

\bibitem[{{Hartmann}(2009)}]{hart09}
---. 2009, {Accretion Processes in Star Formation: Second Edition} ({Cambridge
  University Press})

\bibitem[{{Hartmann} \& {Kenyon}(1996)}]{hartken96}
{Hartmann}, L., \& {Kenyon}, S.~J. 1996, \araa, 34, 207

\bibitem[{{Hatchell} {et~al.}(2007{\natexlab{a}}){Hatchell}, {Fuller}, \&
  {Richer}}]{hatchell07}
{Hatchell}, J., {Fuller}, G.~A., \& {Richer}, J.~S. 2007{\natexlab{a}}, \aap,
  472, 187

\bibitem[{{Hatchell} {et~al.}(2007{\natexlab{b}}){Hatchell}, {Fuller},
  {Richer}, {Harries}, \& {Ladd}}]{hatchell07b}
{Hatchell}, J., {Fuller}, G.~A., {Richer}, J.~S., {Harries}, T.~J., \& {Ladd},
  E.~F. 2007{\natexlab{b}}, \aap, 468, 1009

\bibitem[{{Herbig}(1977)}]{herbig77}
{Herbig}, G.~H. 1977, \apj, 217, 693

\bibitem[{{Herczeg} \& {Hillenbrand}(2008)}]{hh08}
{Herczeg}, G.~J., \& {Hillenbrand}, L.~A. 2008, \apj, 681, 594

\bibitem[{{Hollenbach} {et~al.}(1989){Hollenbach}, {Chernoff}, \&
  {McKee}}]{hollen89}
{Hollenbach}, D.~J., {Chernoff}, D.~F., \& {McKee}, C.~F. 1989, in ESA Special
  Publication, Vol. 290, Infrared Spectroscopy in Astronomy, ed.
  {E.~B{\"o}hm-Vitense}, 245--258

\bibitem[{{Karska} {et~al.}(2013){Karska}, {Herczeg}, {van Dishoeck},
  {Wampfler}, {Kristensen}, {Goicoechea}, {Visser}, {Nisini}, {San
  Jos{\'e}-Garc{\'{\i}}a}, {Bruderer}, {{\'S}niady}, {Doty}, {Fedele},
  {Y{\i}ld{\i}z}, {Benz}, {Bergin}, {Caselli}, {Herpin}, {Hogerheijde},
  {Johnstone}, {J{\o}rgensen}, {Liseau}, {Tafalla}, {van der Tak}, \&
  {Wyrowski}}]{karska13}
{Karska}, A., {Herczeg}, G.~J., {van Dishoeck}, E.~F., {et~al.} 2013, \aap,
  552, A141

\bibitem[{{Kenyon}(1995)}]{kenyon95}
{Kenyon}, S.~J. 1995, in Revista Mexicana de Astronomia y Astrofisica, vol. 27,
  Vol.~1, Revista Mexicana de Astronomia y Astrofisica Conference Series, ed.
  S.~{Lizano} \& J.~M. {Torrelles}, 237

\bibitem[{{Kenyon} {et~al.}(1990){Kenyon}, {Hartmann}, {Strom}, \&
  {Strom}}]{ken90}
{Kenyon}, S.~J., {Hartmann}, L.~W., {Strom}, K.~M., \& {Strom}, S.~E. 1990,
  \aj, 99, 869

\bibitem[{{Kim} {et~al.}(2008){Kim}, {Hirota}, {Honma}, {Kobayashi},
  {Bushimata}, {Choi}, {Imai}, {Iwadate}, {Jike}, {Kameno}, {Kameya},
  {Kamohara}, {Kan-Ya}, {Kawaguchi}, {Kuji}, {Kurayama}, {Manabe}, {Matsui},
  {Matsumoto}, {Miyaji}, {Nagayama}, {Nakagawa}, {Oh}, {Omodaka}, {Oyama},
  {Sakai}, {Sasao}, {Sato}, {Sato}, {Shibata}, {Tamura}, \&
  {Yamashita}}]{kim08}
{Kim}, M.~K., {Hirota}, T., {Honma}, M., {et~al.} 2008, \pasj, 60, 991

\bibitem[{{Krause}(2010)}]{Krause10}
{Krause}, O. 2010, in From Stars to Galaxies: Connecting our Understanding of
  Star and Galaxy Formation, 21

\bibitem[{{Lada}(1985)}]{lada85}
{Lada}, C.~J. 1985, \araa, 23, 267

\bibitem[{{Launhardt} {et~al.}(2013){Launhardt}, {Stutz}, {Schmiedeke},
  {Henning}, {Krause}, {Balog}, {Beuther}, {Birkmann}, {Hennemann},
  {Kainulainen}, {Khanzadyan}, {Linz}, {Lippok}, {Nielbock}, {Pitann}, {Ragan},
  {Risacher}, {Schmalzl}, {Shirley}, {Stecklum}, {Steinacker}, \&
  {Tackenberg}}]{laun13}
{Launhardt}, R., {Stutz}, A.~M., {Schmiedeke}, A., {et~al.} 2013, \aap, 551,
  A98

\bibitem[{{Lee} {et~al.}(2014{\natexlab{a}}){Lee}, {Lee}, {Lee}, {Evans}, \&
  {Green}}]{lee14b}
{Lee}, J.-E., {Lee}, J., {Lee}, S., {Evans}, II, N.~J., \& {Green}, J.~D.
  2014{\natexlab{a}}, \apjs, 214, 21

\bibitem[{Lee(2013)}]{nada}
Lee, L. 2013, NADA: Nondetects And Data Analysis for environmental data, r
  package version 1.5-6

\bibitem[{{Lee} {et~al.}(2014{\natexlab{b}}){Lee}, {Lee}, {Bergin}, \&
  {Park}}]{lee14}
{Lee}, S., {Lee}, J.-E., {Bergin}, E.~A., \& {Park}, Y.-S. 2014{\natexlab{b}},
  \apjs, 213, 33

\bibitem[{{Machida} \& {Hosokawa}(2013)}]{machida13}
{Machida}, M.~N., \& {Hosokawa}, T. 2013, \mnras, 431, 1719

\bibitem[{{Manoj} {et~al.}(2013){Manoj}, {Watson}, {Neufeld}, {Megeath},
  {Vavrek}, {Yu}, {Visser}, {Bergin}, {Fischer}, {Tobin}, {Stutz}, {Ali},
  {Wilson}, {Di Francesco}, {Osorio}, {Maret}, \& {Poteet}}]{manoj13}
{Manoj}, P., {Watson}, D.~M., {Neufeld}, D.~A., {et~al.} 2013, \apj, 763, 83

\bibitem[{{Matt} \& {Pudritz}(2005)}]{mp05}
{Matt}, S., \& {Pudritz}, R.~E. 2005, \apjl, 632, L135

\bibitem[{{Matt} \& {Pudritz}(2008)}]{mp08}
---. 2008, \apj, 681, 391

\bibitem[{{Matuszak} {et~al.}(2015){Matuszak}, {Karska}, {Kristensen},
  {Herczeg}, {Tychoniec}, {van Kempen}, \& {Fuente}}]{matuszak15}
{Matuszak}, M., {Karska}, A., {Kristensen}, L.~E., {et~al.} 2015, ArXiv
  e-prints, arXiv:1504.03347

\bibitem[{{Meeus} {et~al.}(2013){Meeus}, {Salyk}, {Bruderer}, {Fedele},
  {Maaskant}, {Evans}, {van Dishoeck}, {Montesinos}, {Herczeg}, {Bouwman},
  {Green}, {Dominik}, {Henning}, \& {Vicente}}]{Meeus13}
{Meeus}, G., {Salyk}, C., {Bruderer}, S., {et~al.} 2013, \aap, 559, A84

\bibitem[{{Menten} {et~al.}(2007){Menten}, {Reid}, {Forbrich}, \&
  {Brunthaler}}]{menten07}
{Menten}, K.~M., {Reid}, M.~J., {Forbrich}, J., \& {Brunthaler}, A. 2007, \aap,
  474, 515

\bibitem[{{Muzerolle} {et~al.}(1998{\natexlab{a}}){Muzerolle}, {Calvet}, \&
  {Hartmann}}]{mch98}
{Muzerolle}, J., {Calvet}, N., \& {Hartmann}, L. 1998{\natexlab{a}}, \apj, 492,
  743

\bibitem[{{Muzerolle} {et~al.}(2001){Muzerolle}, {Calvet}, \&
  {Hartmann}}]{muzerolle01}
---. 2001, \apj, 550, 944

\bibitem[{{Muzerolle} {et~al.}(2004){Muzerolle}, {D'Alessio}, {Calvet}, \&
  {Hartmann}}]{muz04}
{Muzerolle}, J., {D'Alessio}, P., {Calvet}, N., \& {Hartmann}, L. 2004, \apj,
  617, 406

\bibitem[{{Muzerolle} {et~al.}(1998{\natexlab{b}}){Muzerolle}, {Hartmann}, \&
  {Calvet}}]{muz98}
{Muzerolle}, J., {Hartmann}, L., \& {Calvet}, N. 1998{\natexlab{b}}, \aj, 116,
  2965

\bibitem[{{Myers} \& {Ladd}(1993)}]{ml93}
{Myers}, P.~C., \& {Ladd}, E.~F. 1993, \apjl, 413, L47

\bibitem[{{Najita} \& {Shu}(1994)}]{ns94}
{Najita}, J.~R., \& {Shu}, F.~H. 1994, \apj, 429, 808

\bibitem[{{Nisini} {et~al.}(2015){Nisini}, {Santangelo}, {Giannini},
  {Antoniucci}, {Cabrit}, {Codella}, {Davis}, {Eisl{\"o}ffel}, {Kristensen},
  {Herczeg}, {Neufeld}, \& {van Dishoeck}}]{nisini15}
{Nisini}, B., {Santangelo}, G., {Giannini}, T., {et~al.} 2015, \apj, 801, 121

\bibitem[{{Pelletier} \& {Pudritz}(1992)}]{pp92}
{Pelletier}, G., \& {Pudritz}, R.~E. 1992, \apj, 394, 117

\bibitem[{{Plunkett} {et~al.}(2015){Plunkett}, {Arce}, {Corder}, {Dunham},
  {Garay}, \& {Mardones}}]{plunkett15b}
{Plunkett}, A.~L., {Arce}, H.~G., {Corder}, S.~A., {et~al.} 2015, \apj, 803, 22

\bibitem[{{R Core Team}(2015)}]{Rpro}
{R Core Team}. 2015, R: A Language and Environment for Statistical Computing, R
  Foundation for Statistical Computing, Vienna, Austria

\bibitem[{{Reipurth} \& {Aspin}(2010)}]{ra10}
{Reipurth}, B., \& {Aspin}, C. 2010, in Evolution of Cosmic Objects through
  their Physical Activity, ed. H.~A. {Harutyunian}, A.~M. {Mickaelian}, \&
  Y.~{Terzian}, 19--38

\bibitem[{{Richer} {et~al.}(2000){Richer}, {Shepherd}, {Cabrit}, {Bachiller},
  \& {Churchwell}}]{richer00}
{Richer}, J.~S., {Shepherd}, D.~S., {Cabrit}, S., {Bachiller}, R., \&
  {Churchwell}, E. 2000, Protostars and Planets IV, 867

\bibitem[{{Rodriguez} {et~al.}(1982){Rodriguez}, {Carral}, {Ho}, \&
  {Moran}}]{rodri82}
{Rodriguez}, L.~F., {Carral}, P., {Ho}, P.~T.~P., \& {Moran}, J.~M. 1982, \apj,
  260, 635

\bibitem[{{Safron} {et~al.}(2015){Safron}, {Fischer}, {Megeath}, {Furlan},
  {Stutz}, {Stanke}, {Billot}, {Rebull}, {Tobin}, {Ali}, {Allen}, {Booker},
  {Watson}, \& {Wilson}}]{safron15}
{Safron}, E.~J., {Fischer}, W.~J., {Megeath}, S.~T., {et~al.} 2015, \apjl, 800,
  L5

\bibitem[{{Sandstrom} {et~al.}(2007){Sandstrom}, {Peek}, {Bower}, {Bolatto}, \&
  {Plambeck}}]{sandstrom07}
{Sandstrom}, K.~M., {Peek}, J.~E.~G., {Bower}, G.~C., {Bolatto}, A.~D., \&
  {Plambeck}, R.~L. 2007, \apj, 667, 1161

\bibitem[{{Sen}(1968)}]{sen68}
{Sen}, P.~K. 1968, Journal of the American Statistical Association, 63, 1379

\bibitem[{{Shu} {et~al.}(1994){Shu}, {Najita}, {Ostriker}, {Wilkin}, {Ruden},
  \& {Lizano}}]{shu94}
{Shu}, F., {Najita}, J., {Ostriker}, E., {et~al.} 1994, \apj, 429, 781

\bibitem[{{Snell}(1987)}]{snell87}
{Snell}, R.~L. 1987, in IAU Symposium, Vol. 115, Star Forming Regions, ed.
  M.~{Peimbert} \& J.~{Jugaku}, 213--236

\bibitem[{{Stanke} {et~al.}(2010){Stanke}, {Stutz}, {Tobin}, {Ali}, {Megeath},
  {Krause}, {Linz}, {Allen}, {Bergin}, {Calvet}, {di Francesco}, {Fischer},
  {Furlan}, {Hartmann}, {Henning}, {Manoj}, {Maret}, {Muzerolle}, {Myers},
  {Neufeld}, {Osorio}, {Pontoppidan}, {Poteet}, {Watson}, \&
  {Wilson}}]{stanke10}
{Stanke}, T., {Stutz}, A.~M., {Tobin}, J.~J., {et~al.} 2010, \aap, 518, L94

\bibitem[{{Sturm} {et~al.}(2013){Sturm}, {Bouwman}, {Henning}, {Evans},
  {Waters}, {van Dishoeck}, {Green}, {Olofsson}, {Meeus}, {Maaskant},
  {Dominik}, {Augereau}, {Mulders}, {Acke}, {Merin}, \& {Herczeg}}]{sturm13}
{Sturm}, B., {Bouwman}, J., {Henning}, T., {et~al.} 2013, \aap, 553, A5

\bibitem[{{Stutz} {et~al.}(2010){Stutz}, {Launhardt}, {Linz}, {Krause},
  {Henning}, {Kainulainen}, {Nielbock}, {Steinacker}, \& {Andr{\'e}}}]{stutz10}
{Stutz}, A., {Launhardt}, R., {Linz}, H., {et~al.} 2010, \aap, 518, L87

\bibitem[{{Stutz} \& {Kainulainen}(2015)}]{stutz15}
{Stutz}, A.~M., \& {Kainulainen}, J. 2015, \aap, 577, L6

\bibitem[{{Stutz} {et~al.}(2013){Stutz}, {Tobin}, {Stanke}, {Megeath},
  {Fischer}, {Robitaille}, {Henning}, {Ali}, {di Francesco}, {Furlan},
  {Hartmann}, {Osorio}, {Wilson}, {Allen}, {Krause}, \& {Manoj}}]{stutz13}
{Stutz}, A.~M., {Tobin}, J.~J., {Stanke}, T., {et~al.} 2013, \apj, 767, 36

\bibitem[{{Takahashi} {et~al.}(2008){Takahashi}, {Saito}, {Ohashi}, {Kusakabe},
  {Takakuwa}, {Shimajiri}, {Tamura}, \& {Kawabe}}]{takahashi08}
{Takahashi}, S., {Saito}, M., {Ohashi}, N., {et~al.} 2008, \apj, 688, 344

\bibitem[{{Tobin}(2010)}]{tobin10}
{Tobin}, J. 2010, {OT1\_jtobin\_1: Protostellar Envelopes Resolved Inside and
  Out: A Close Look in the Far-IR}, Herschel Space Observatory Proposal,
  id.752, ,

\bibitem[{{Tobin} {et~al.}(2012){Tobin}, {Hartmann}, {Chiang}, {Wilner},
  {Looney}, {Loinard}, {Calvet}, \& {D'Alessio}}]{tobin12}
{Tobin}, J.~J., {Hartmann}, L., {Chiang}, H.-F., {et~al.} 2012, \nat, 492, 83

\bibitem[{{Tobin} {et~al.}(2015){Tobin}, {Stutz}, {Megeath}, {Fischer},
  {Henning}, {Ragan}, {Ali}, {Stanke}, {Manoj}, {Calvet}, \&
  {Hartmann}}]{tobin15}
{Tobin}, J.~J., {Stutz}, A.~M., {Megeath}, S.~T., {et~al.} 2015, \apj, 798, 128

\bibitem[{{Tobin} {et~al.}(2016){Tobin}, {Stutz}, {Manoj}, {Megeath}, {Karska},
  {Nagy}, {Wyrowski}, {Fischer}, {Watson}, \& {Stanke}}]{tobin16}
{Tobin}, J.~J., {Stutz}, A.~M., {Manoj}, P., {et~al.} 2016, ArXiv e-prints,
  arXiv:1607.00787

\bibitem[{{van Kempen} {et~al.}(2010){van Kempen}, {Green}, {Evans}, {van
  Dishoeck}, {Kristensen}, {Herczeg}, {Mer{\'{\i}}n}, {Lee}, {J{\o}rgensen},
  {Bouwman}, {Acke}, {Adamkovics}, {Augereau}, {Bergin}, {Blake}, {Brown},
  {Carr}, {Chen}, {Cieza}, {Dominik}, {Dullemond}, {Dunham}, {Glassgold},
  {G{\"u}del}, {Harvey}, {Henning}, {Hogerheijde}, {Jaffe}, {Kim}, {Knez},
  {Lacy}, {Maret}, {Meeus}, {Meijerink}, {Mulders}, {Mundy}, {Najita},
  {Olofsson}, {Pontoppidan}, {Salyk}, {Sturm}, {Visser}, {Waters}, {Waelkens},
  \& {Y{\i}ld{\i}z}}]{vankemp10b}
{van Kempen}, T.~A., {Green}, J.~D., {Evans}, N.~J., {et~al.} 2010, \aap, 518,
  L128

\bibitem[{{Vorobyov} \& {Basu}(2008)}]{vb08}
{Vorobyov}, E.~I., \& {Basu}, S. 2008, \apjl, 676, L139

\bibitem[{{Vorobyov} \& {Basu}(2010)}]{vb10}
---. 2010, \apj, 719, 1896

\bibitem[{{Wardle} \& {Koenigl}(1993)}]{wk93}
{Wardle}, M., \& {Koenigl}, A. 1993, \apj, 410, 218

\bibitem[{{Watson}(1985)}]{watson85}
{Watson}, D.~M. 1985, Physica Scripta Volume T, 11, 33

\bibitem[{{Wu} {et~al.}(2004){Wu}, {Wei}, {Zhao}, {Shi}, {Yu}, {Qin}, \&
  {Huang}}]{wu04}
{Wu}, Y., {Wei}, Y., {Zhao}, M., {et~al.} 2004, \aap, 426, 503

\end{thebibliography}

\end{document}